\documentclass[12pt]{article}
\usepackage{amsfonts}
\usepackage{mathrsfs}
\usepackage{amsthm}
\usepackage{multirow}
\usepackage{bbding}
\usepackage{amssymb}
\usepackage{amsmath}
\usepackage{graphicx,color}
\usepackage{rotating}

\usepackage{bm}
\usepackage{epstopdf}
\DeclareMathOperator*{\argmin}{arg\,min}

\newcommand{\bea}{\begin{eqnarray}}
\newcommand{\eea}{\end{eqnarray}}
\newcommand{\Bea}{\begin{eqnarray*}}
\newcommand{\Eea}{\end{eqnarray*}}
\newcommand{\ba}{\begin{array}}
\newcommand{\ea}{\end{array}}
\newcommand{\bt}{\begin{tabular}}
\newcommand{\et}{\end{tabular}}
\newcommand{\btb}{\begin{table}}
\newcommand{\etb}{\end{table}}
\newcommand{\bc}{\begin{center}}
\newcommand{\ec}{\end{center}}
\newcommand{\beq}{\begin{equation}}
\newcommand{\eeq}{\end{equation}}

\usepackage{float}

\makeatletter

\newcommand{\Rmnum}[1]{\expandafter\@slowromancap\romannumeral #1@}
\makeatother

\begin{document}

\title{A race-DC in Big Data }
\author{Lu Lin$^1$, and Jun Lu$^2$\footnote{The corresponding
author. Email: lujun@zjgsu.edu.cn. The research was
supported by NNSF projects (11971265) of China.}
\\
\small $^1$Zhongtai Securities Institute for Financial Studies, Shandong University, Jinan, China\\
\small $^2$School of Statistics and Mathematics, Zhejiang Gongshang University, Hangzhou, China
 }
\date{}
\maketitle

\vspace{-0.3cm}

\begin{abstract} \baselineskip=18pt

The strategy of divide-and-combine (DC) has been widely used in the area of big data.
Bias-correction is crucial in the DC procedure for validly aggregating the locally biased estimators, especial for the case when the number of batches of data is large.
This paper establishes a race-DC through a residual-adjustment composition estimate (race).
The race-DC applies to various types of biased estimators, which include but are not limited to Lasso estimator, Ridge estimator and principal component estimator in linear regression, and least squares estimator in nonlinear regression. The resulting global estimator is strictly unbiased under linear model, and is acceleratingly bias-reduced in nonlinear model, and can achieve the theoretical optimality, for the case when the number of batches of data is large. Moreover, the race-DC is computationally simple because it is a least squares estimator in a pro forma linear regression.
Detailed simulation studies demonstrate
that the resulting global estimator is significantly bias-corrected, and the behavior is
comparable with the oracle estimation and is much better than the competitors.


\end{abstract}

\baselineskip=20pt

\newpage

\setcounter{equation}{0}
\section{Introduction}

The modern developments of science and technology have enabled the massive data sets arising in various fields. The major challenge in analyzing this kind of data is that data storage and analysis by a single computer are hardly feasible. The divide-and-conquer (DC) is the most commonly used approach in computer science to deal with the massive datasets. Typically, a DC algorithm first distributes the massive datasets to many machines with limited memory, and then receives the locally compressed data delivered by each machine, and finally, aggregates these compressed data to achieve the global results (e.g., global parameter estimator).
To validly compress the raw data and efficiently aggregate the compressed data in the DC procedure, various types of methodologies have been proposed in the statistical literature. These methods include but are not limited to the naive average of the local estimators (Mcdonald et al. 2009; Zinkevich et al. 2010), the relevant DC expressions (Chen, et al., 2006; Lin and Xi, 2011) and the representative approaches (Li and Yang, 2018; Wang, et al., 2019).
Moreover, these methods have been widely applied to various statistical models, such as parameter models (Chen et al., 2006; Lin and Xi, 2011; Schifano et al., 2016; Chen and Xie, 2014; Zhang, Duchi and Wainwright, 2015), high-dimensional parametric regressions (Lee et al., 2017; Battey et al., 2018), semi-parametric regressions (Zhao, Cheng and Liu, 2016), quantile regressions (Chen, Liu and Zhang, 2018; Volgushev, Chao and Cheng, 2019), nonlinear models (Shi, Lu and Song, 2018), and nonparametric models (Li, Lin and Li, 2013).

However, the estimation bias among most of the methods aforementioned cannot be sufficiently reduced, especially when the number of batches of data (or machines) is large.
In order to achieve the same efficiency as that of the oracle estimator with access to entire datasets, some special data processing technologies were proposed,
see the local bias-corrections of Lee et al. (2017), Lian et al. (2018) and Keren and Yang (2018), and the iterative algorithms of Jordan, Lee and Yang (2019), Wang et al. (2017), and  Chen, Liu and Zhang (2018). Sometimes these methodologies are computationally complicated, and furthermore, the relevant bias-corrections are not efficient such that some constraints on the number of batches of data are still imposed, that is to say, the number of batches of data should be relatively small. This restricts the application of these methods in some situations where the number of batches could be very large. For example, in a large-scale sensor network, the number of sensors could be very large but each sensor can only collect and store a limited amount of data. Another representative example is the online streaming data, the processor receives lots of batches of data in a short time but the memory of it is usually limited.

Because of wide existence of biased estimation and the important position of bias-correction in the procedure of DC, we in this paper focus on the issue of general biased estimations in both linear and nonlinear regressions under the environment of big data. Instead of designing local bias-correction method and iterative algorithm, we explore a global bias-correction algorithm from a new perspective.
The newly proposed methodology is motivated by the proven data analysis technique --- statistical composition. This method has been studied in the community of machine learning, see for example, Wang et al. (2019), Dai et al. (2016 and 2017), Wang and Lin (2015), and Tong and Wang (2005), in which they suggested some composition methods for constructing various composite estimators of the derivative and variance in nonparametric regression.
Composition method has also attained much attention in the community of statistic. For instance, in quantile regression and nonparametric regression, statisticians have achieved the optimal estimation efficiency via optimizing the composite estimation covariance, see Zou and Yuan (2008); Kai, Li, and Zou (2010); Sun, Gai, and Lin (2013); Kai, Li, and Zou (2011); Bradic, Fan, and Wang (2011) and among others.
Recently, the composition idea has also been utilized to reduce the estimation bias for various statistical models, see Lin et al. (2019); Cheng et al. (2018); Lin and Li (2008) for further reading.

The composition method, however, has not been well investigated in the
area of big data, to the best of our knowledge.
In this paper, to achieve the goal of global bias-correction under the environment of big data, a new DC method is introduced via {\it race} --- residual-adjustment composition estimate. The DC method is then denoted by race-DC because of the use of race. The race-DC first constructs a residual-adjustment estimator based on general biased estimators, and then establishes a pro forma linear model that regards the residual-adjustment estimator as response variable, and the parameter of interest as regression coefficient. Consequently, a global biased-reduced estimator can be obtained by the least squares composition in the pro forma linear model. It will be shown in later sections that our method has the following salient features.

\begin{enumerate}
\item[1)] {\it Generality.} The race-DC applies to various types of biased estimators, including Lasso estimator, Ridge estimator and principal component estimator in linear regression, and least squares estimator in nonlinear regression.
\item[2)] {\it Global bias-correction.} The resulting global
estimator is strictly unbiased in linear model, and is acceleratingly bias-reduced in nonlinear model, even for the case when the local estimators have a non-negligible bias.
\item[3)] {\it Standard convergence rate.} The global estimator achieves the root-consistency, a standard one as pooling all the data together, for the case when the number of batches of data is relatively large.
\item[4)] {\it Simplicity.} The estimation procedure of race-DC is computationally simple. More specifically, the remodelling process is essentially a simple residual-adjustment procedure with a succinct form of linear models, thus the aggregation procedure is just the simple least squares.
\end{enumerate}
These salient features will be confirmed by theoretical properties. Moreover, our simulation studies particularly show that the global estimator is significantly bias-corrected,
and its behavior is much better than the competitors and is comparable with the oracle estimator.

The remainder of this paper is organized in the following way. In Section 2, after recalling the debiased lasso estimator in linear model, a residual-adjustment remodeling is introduced, a least square aggregation is suggested, and the theoretical properties of the resulting global estimator is investigated. In Section 3, the residual-adjustment remodeling and least squares aggregation are extended into the non-linear model. Simulation studies are provided in Section 4 to illustrate
the new method.

\setcounter{equation}{0}
\section{The race-DC in linear model}

\subsection{Motivating example: Lasso estimator}
In this section, we consider the following linear model:
\begin{eqnarray}\label{(model)}Y_i=X_i^T\beta +\varepsilon_i, \ i=1,\cdots,n,\end{eqnarray}  where $\beta=(\beta^1,\cdots,\beta^p)^T$ is a $p$-dimensional vector of unknown parameters, $X_i=(X_i^1,\cdots,X_i^p)^T$, $i=1,\cdots,n$, are independent observations of a $p$-dimensional covariate $X=(X^1,\cdots,X^p)^T$, and the errors $\varepsilon_i,i=1,\cdots,n$, are independent and identically distributed as $\varepsilon$, satisfying ${\rm E}[\varepsilon_i|X_i]=0$ and ${\rm Var}[\varepsilon_i|X_i]=\sigma_\varepsilon^2$. For convenience of modeling, suppose the dimension $p$ is fixed. The compression and aggregation methods proposed below still apply to the case where $p$ depends on $n$.

When the sample size $n$ is extremely large, the index set $\{1,\cdots, n\}$ is split into $N$ subsets ${\cal H}_1,\cdots,{\cal H}_N$ with size $m=|{\cal H}_j|$ satisfying $n= Nm$. Correspondingly, the entire dataset $\{(Y_i,X_i),i=1,\cdots,n\}$ is divided into $N$ batches ${\cal D}_1,\cdots,{\cal D}_N$ with ${\cal D}_j=\{(Y_i,X_i),i\in{\cal H}_j\}$.
According to Tibshirani (1996),
the lasso estimator $\widehat\beta^{Lasso}_j$ of $\beta$ computed on ${\cal D}_j$ is defined as
\begin{eqnarray}\label{(lasso)}\widehat\beta^{Lasso}_j=\argmin_{\beta}\frac{1}{2m}\sum_{i\in{\cal H}_j}(Y_i-X_i^T\beta)^2+\lambda_j\|\beta\|_1,\end{eqnarray} where $\lambda_j>0$ is a regularization parameter.
It is known that the lasso estimator is biased. To reduce the bias of lasso estimator $\widehat\beta^{Lasso}_j$ in (\ref{(lasso)}), Javanmard and Montanari (2014) proposed a debiased lasso estimator, which has the form of
\begin{eqnarray}\label{(debiased-lasso)}\widehat\beta^{DBL}_j=\widehat\beta^{Lasso}_j+\frac1m\widehat{\bm\Theta}_j {\bf X}_j^T({\bf y}_j-{\bf X}_j\widehat\beta^{Lasso}_j),\end{eqnarray} where ${\bf X}_{j}=(X_{l}:l\in{\cal H}_j)^T$, ${\bf y}_j=(y_{l}:l\in{\cal H}_j)^T$, and $\widehat{\bm\Theta}_j$ is an approximate inverse to $\frac1m{\bf X}_j^T{\bf X}_j$. Such an estimator is bias-decayed because the
correction term $\frac1m\widehat{\bm\Theta}_j {\bf X}_j^T({\bf y}_j-{\bf X}_j\widehat\beta^{Lasso}_j)$ is a subgradient of $\lambda_j\|\cdot\|_1$ at $\widehat\beta^{Lasso}_j$.
For general remodeling goal (see Subsection 2.3 and Section 3), in this paper we call the estimator as residual-adjustment lasso estimator. On the other hand, when $\frac1m{\bf X}_j^T{\bf X}_j$ is non-singular, setting $\widehat{\bm\Theta}_j=(\frac1m{\bf X}_j^T{\bf X}_j)^{-1}$ gives the ordinary least squares estimator. For general case, as was shown by Javanmard and Montanari (2014) and Lee et al. (2017), the matrix $\widehat{\bm\Theta}_j$ can be constructed via the method of row by row coherence, but the corresponding implementation procedure is computationally complicated.

Since the estimators $\widehat\beta^{DBL}_j, j=1,\cdots,N$, are bias-reduced, Lee et al. (2017) directly aggregated them by simple average way as
\begin{eqnarray}\label{(average)}\overline\beta^{GDBL}=\frac1m\sum_{j=1}^N\widehat\beta^{DBL}_j.\end{eqnarray}
It was shown by Lee et al. (2017) that the global estimator $\overline\beta^{GDBL}$ can achieve the minimax optimality if $N$ satisfies the condition $N=o(\sqrt n)$ (or equivalently $n=o(m^2)$), which means that $N$ cannot be too large.

In addition to the global estimator $\overline\beta^{GDBL}$, most of global estimators derived from various DC methods also need the constraint:
\begin{eqnarray}\label{(constraint)}N=o(\sqrt n) \ \mbox{or its similar and improved versions},\end{eqnarray}
see, e.g., the global estimator in estimating equation estimation (Lin and Xi, 2011), and the global estimator in quantile regression (Chen, Liu and Zhang, 2018).
However, as mentioned above, the constraint on $N$ heavily restricts its application in many scenarios such as the large-scale sensor networks and the online streaming data.



\subsection{The race-DC for Lasso estimator}
Note that the local estimator $\widehat\beta^{DBL}_j$ is still biased and averaging them only reduces variance, but not bias. As a result, we still need the condition of small $N$ as in (\ref{(constraint)}) to achieve the minimax optimality. As mentioned above, however, the number of batches $N$ could be very large in some applications which would make the local bias-correction and iterative algorithm methods invalid. To solve this problem, we in this subsection develop a new bias-correction method specially for the case of large $N$. To this end, the following condition is required:
\begin{enumerate}
\item[{\it C1}.] $N>p$.\end{enumerate}

By model (\ref{(model)}) and the formula of the residual-adjustment lasso estimator $\widehat\beta^{DBL}_j$ in (\ref{(debiased-lasso)}),
we can rewrite the residual-adjustment lasso estimator as
\begin{eqnarray}\label{(rewritten)}\widehat\beta^{DBL}_j=\widehat\beta^{Lasso}_j+\frac1m\widehat{\bm\Theta}_j {\bf X}_j^T{\bf X}_j(\beta-\widehat\beta^{Lasso}_j)+\widehat{\bm\epsilon}_j,\end{eqnarray}
where $\widehat{\bm\epsilon}_j=\frac1m\widehat{\bm\Theta}_j {\bf X}_j^T\bm\varepsilon_j$ with $\bm\varepsilon_j=(\varepsilon_i:i\in{\cal H}_j)^T$. Let $\eta_j$ be a $p$-dimensional random vector (the choice of $\eta_j$ will be given later), then we get the following pro form linear regression model:
\begin{eqnarray}
\label{(remodeling)}\widehat{z}^{Lasso}_j=\widehat{U}^T_j\beta+\widehat{\epsilon}_j,\ j=1,\cdots,N,\end{eqnarray}
where $\widehat{U}_j=\frac1m {\bf X}_j^T{\bf X}_j\widehat{\bm\Theta}^T_j\eta_j$ is $p$-dimensional covariate, $\widehat{z}^{Lasso}_j=\eta^T_j(\widehat\beta^{DBL}_j-\widehat\beta^{Lasso}_j)
+\widehat{U}^T_j\widehat\beta^{Lasso}_j$ is a scalar response variable, $\widehat{\epsilon}_j=\eta^T_j\frac1m\widehat{\bm\Theta}_j {\bf X}_j^T\bm\varepsilon_j$ is the error term satisfying
\begin{eqnarray}\label{(error)}E[\widehat{\epsilon}_j|{\bf X}_j]=0\ \mbox{ and } \ Var[\widehat{\epsilon}_j|{\bf X}_j]=\frac{\sigma_\varepsilon^2}{m}\widehat\sigma^2_j\end{eqnarray} with $\widehat\sigma^2_j=\frac{1}{m}\eta^T_j\widehat{\bm\Theta}_j {\bf X}_j^T{\bf X}_j\widehat{\bm\Theta}_j^T\eta_j$.

\

{\bf Remark 2.1.} {\it Model (\ref{(remodeling)}) has the following features:
\begin{enumerate}
\item[(1)] Under condition (C1), the size $N$ of ``dataset" $\{(\widehat{z}^{Lasso}_j,\widehat{U}^T_j):j=1,\cdots,N\}$ in model (\ref{(remodeling)}) is large so that a good estimator of $\beta$ is available by the least square. However, the original linear model (\ref{(model)}) confined on dataset ${\cal D}_j$ might be invalid because the size $m$ of ${\cal D}_j$ may not be large enough to result in a good estimator.
\item[(2)] The main difference from the standard regression is that the variances of errors $\widehat{\epsilon}_j$ are of order $O(1/m)$ as in (\ref{(error)}), rather than a constant. This is a key for achieving the standard convergence rate; for the details see Theorem 2.1 and Remark 2.2 given below. 
\item[(3)] Linear regression model (\ref{(remodeling)}) is actually the random projection of the representation (\ref{(rewritten)}) of debiased lasso estimator. It is well known that random projection of appropriate sizes preserves enough information for exact reconstruction with high probability, specially for least squares estimation in linear model (see, e.g., Fard, et al., 2012).

\item[(4)] Although (\ref{(remodeling)}) is a type of projection model, actually this special form is derived from the residual-adjustment structure as in (\ref{(debiased-lasso)}). On the other hand, by this treatment, we can choose suitable inverse matrices $\widehat{\bm\Theta}_j$ such that the resulting global estimator can return to least squares estimator when $\frac1m{\bf X}_j^T{\bf X}_j,j=1,\cdots,N$, are non-singular or nearly non-singular (see Remark 2.2) and still works well even when the least squares is inapplicable in some sub-datasets (e.g., there is colinearity between some predictors in some sub-datasets). Therefore, the term ``residual-adjustment", instead of projection, will be used in the following, and the residual-adjustment method will be further extended into general case of bias estimation later.
\end{enumerate}}

Then, in the following, we use model (\ref{(remodeling)}) to construct global estimator.
For establishing a valid regression, we need the following condition:
\begin{enumerate}
\item[{\it C2}.] The matrix $\frac1N\sum_{j=1}^N\widehat\sigma_j^{-2}\widehat{U}_j\widehat{U}_j^T$ is positive definite, where $\widehat\sigma_j^{-2}$ is an approximate inverse to $\widehat\sigma_j^2$.\end{enumerate}
Because of {\it C1}, Condition {\it C2} is common if $\eta_j,j=1,\cdots,N$, are independent and identically distributed as a random vector $\eta\sim F_\eta$ with given distribution $F_\eta$. For example, consider the case where the approximate inverse is chosen as $\widehat\sigma_j^{-2}=1$ for all $j$, and $\widehat{\bm\Theta}_j\rightarrow_p{\bm\Theta}$ as $m\rightarrow
\infty$ for all $j$ (see the example given in Remark 2.3). Let $\Sigma_X$ and $\Sigma_\eta$ be the covariance matrices of $X$ and $\eta$, respectively. In this case, $\frac1N\sum_{j=1}^N\widehat\sigma_j^{-2}\widehat{U}_j\widehat{U}_j^T
\approx \Sigma_X{\bm\Theta}^T\Sigma_\eta{\bm\Theta}\Sigma_X$, which is positive definite provided that both ${\bm\Theta}$ and $\Sigma_\eta$ are positive definite.

Combining the linear regression (\ref{(remodeling)}) together with the expectation and variance conditions in (\ref{(error)}), we aggregate $\widehat\beta^{DBL}_j$ and $\widehat\beta^{Lasso}_j$ by solving the following weighted least squares:
$$\sum_{j=1}^N\widehat\sigma_j^{-2}(\widehat{z}^{Lasso}_j-\widehat{U}^T_j\beta)^2.$$
Then,  the global estimator of $\beta$ is available as
\begin{eqnarray}\label{(GUE)}\widetilde\beta^{GRAL}=
\left(\sum_{j=1}^N\widehat\sigma_j^{-2}\widehat{U}_j\widehat{U}_j^T\right)^{-1}\sum_{j=1}^N
\widehat\sigma_j^{-2}\widehat{U}_j\widehat{z}^{Lasso}_j.\end{eqnarray}

{\bf Remark 2.2.} {\it
\begin{enumerate}
\item[(1)]
Particularly, when $\frac1m{\bf X}_j^T{\bf X}_j,j=1,\cdots,N$, are non-singular (or nearly non-singular) and $\widehat{\bm\Theta}_j$ are chosen as $(\frac1m{\bf X}_j^T{\bf X}_j)^{-1}$ (or its Ridge form; see Remark 2.3), $\widetilde\beta^{GRAL}$ is exactly (or nearly) the ordinary least square estimator constructed by entire datasets.
\item[(2)]
The newly defined estimator $\widetilde\beta^{GRAL}$ in (\ref{(GUE)}) is indeed a DC expression, without accessing any raw data.
It can be seen that the DC method is a residual-adjustment composite estimator (race) via the pro form regression model (\ref{(remodeling)}). Thus, the method is called as {\it race-DC},
and the corresponding estimator $\widetilde\beta^{GRAL}$ is named as global residual-adjustment lasso estimator (GRAL estimator, for short).
\item[(3)] It will be shown in next subsection that in our estimation procedure, precise choices of $\widehat{\bm\Theta}_j$ and $\widehat\sigma_j^{-2}$ are not needed. Actually, they can be constructed by some empirical methods; the details will be given in next subsection and the simulation studies. Therefore, the estimation procedure is computationally simple, without need of any iterative algorithm.
\end{enumerate}}

Unlike sparse lasso estimator, however, the GRAL estimator is usually
dense. The denseness detracts from the intrepretability of the regression coefficients and makes the
estimation error large in the $L_2$ and $L_1$ norms. To remedy the two problems, we threshold the GRAL estimator $\widetilde\beta^{GRAL}$ by the following hard and soft-thresholdings:
\begin{eqnarray}\label{(threshold)}\nonumber&&\widetilde\beta^{ht}=\left(\widetilde\beta^1{\bf 1}_{\{|\widetilde\beta^1|>t\}},\cdots,\widetilde\beta^p{\bf 1}_{\{|\widetilde\beta^p|>t\}}\right)^T,\\&&
\widetilde\beta^{st}=\left(\mbox{sgn}(\widetilde\beta^1)\max\{|\widetilde\beta^1|-t,0\},\cdots,
\mbox{sgn}(\widetilde\beta^p)\max\{|\widetilde\beta^p|-t,0\}\right)^T,\end{eqnarray} where $\widetilde\beta^k$ is the $k$-th component of $\widetilde\beta^{GRAL}$.
The above thresholding GRAL estimators produce sparse aggregations, and possess the same asymptotic properties as those of $\widetilde\beta^{GRAL}$ when $t$ tends to zero at a suitable rate such as $t=O\left(\sqrt{\log p/n}\right)$ (see, e.g., Theorem 17 and Remark 13 of Lee et al., 2017).

To establish the theoretical properties, we first rewrite $\widetilde\beta^{GRAL}$ as
\begin{eqnarray}\label{(GUE-1)}\widetilde\beta^{GRAL}=\beta+
\left(\frac1N\sum_{j=1}^N\widehat\sigma_j^{-2}\widehat{U}_j\widehat{U}_j^T\right)^{-1}\frac1N\sum_{j=1}^N
\widehat\sigma_j^{-2}
\widehat{U}_j\widehat{\epsilon}_j.\end{eqnarray} Note that the errors $\widehat{\epsilon}_j$ have expectation zero and variance of order $O(1/m)$ as in (\ref{(error)}). Then, the following theorem follows directly.

\noindent{\bf Theorem 2.1.} {\it Under conditions C1 and C2, and for any choices of $\widehat{\bm\Theta}_j$ and $\widehat\sigma_j^{-2}$, the GRAL estimator $\widetilde\beta^{GRAL}$ in (\ref{(GUE)}) satisfies
\begin{eqnarray}\label{(bias-variance)}E[\widetilde\beta^{GRAL}|{\bf X}]=\beta \ \mbox{ and } \ Cov[\widetilde\beta^{GRAL}|{\bf X}]=\frac{\sigma_\varepsilon^2}{n}
\left(\frac1N\sum_{j=1}^N\widehat\sigma_j^{-2}\widehat{U}_j\widehat{U}_j^T\right)^{-1},\end{eqnarray}
where ${\bf X}=(X_1,\cdots,X_n)^T$.}

From the theorem, we have the following findings:

{\bf Remark 2.3.} {\it 
\begin{enumerate}
\item[(1)]
The theorem ensures that for the case of large $N$, the GRAL estimator is always unbiased, and the convergence rate in probability is of order $O(1/\sqrt n)$, the standard one of the oracle estimator constructed by entire datasets. The existing methods
cannot achieve this goal.
\item[(2)] The theorem shows that the unbiasedness and convergence rate of the GRAL estimator are basically free of the choices of $\widehat{\bm\Theta}_j$ and $\widehat\sigma_j^{-2}$. For simplifying the procedure of estimation, $\widehat{\bm\Theta}_j$ and $\widehat\sigma_j^{-2}$ can be chosen by empirical methods, for example, the Ridge estimation forms as
\begin{eqnarray*}&&\widehat{\bm\Theta}_j(k_1)=\left(\frac1m{\bf X}_j^T{\bf X}_j+k_1I\right)^{-1}, \\&& \widehat\sigma_j^{-2}(k_1,k_2)=\left(\eta^T_j\widehat{\bm\Theta}_j(k_1) \left(\frac1m{\bf X}_j^T{\bf X}_j+k_2I\right)\widehat{\bm\Theta}_j(k_1)\eta_j\right)^{-1}\end{eqnarray*} for some constants $k_1,k_2>0$.
These show that the estimation method of $\widetilde\beta^{GRAL}$ is computationally simpler than the local bias-reduction and iterative algorithm methods.
\item[(3)] As was stated above, $\eta_j,j=1,\cdots,N$, are independently generated from a given distribution $F_\eta$. According to the existing literature (see, e.g., Fard, et al., 2012), $F_\eta$ can be chosen as a normal distribution with mean zero and variance $1/p$. In the following, we provide a theoretical way on the choice of $F_\eta$. From the definition of $\widehat{{\bm\Theta}}_j$, we have
$$Cov[\widetilde\beta^{GRAL}|{\bf X}]\approx\frac{\sigma_\varepsilon^2}{n}
\left(\frac1N\sum_{j=1}^N\widehat\sigma_j^{-2}\eta_j\eta_j^T\right)^{-1}.$$
We then use  $\mbox{tr}\left(\frac1N\sum_{j=1}^N\widehat\sigma_j^{-2}\eta_j\eta_j^T\right)$ as a criterion for choosing $\eta_j$. Without loss of generality, suppose $\|\eta_j\|_2=1$ for all $j$.
As a result, $\mbox{tr}\left(\frac1N\sum_{j=1}^N\widehat\sigma_j^{-2}\eta_j\eta_j^T\right)
=\mbox{tr}\left(\frac1N\sum_{j=1}^N\widehat\sigma_j^{-2}\right)$. Then, maximizing $\mbox{tr}\left(\frac1N\sum_{j=1}^N\widehat\sigma_j^{-2}\eta_j\eta_j^T\right)$  is equivalent to minimizing
$$\widehat\sigma_j^2(k_1,k_2)=\eta^T_j\widehat{\bm\Theta}_j(k_1) \left(\frac1m{\bf X}_j^T{\bf X}_j+kI_2\right)\widehat{\bm\Theta}_j(k_1)\eta_j.$$ It shows that the optimal choice of $\eta_j$ is the minimum eigenvector of
the matrix $\frac1m\widehat{\bm\Theta}_j(k_1) \left(\frac1m{\bf X}_j^T{\bf X}_j+k_2I\right)\widehat{\bm\Theta}_j(k_1)$. Suppose that
$$\frac1m\widehat{\bm\Theta}_j(k_1) \left(\frac1m{\bf X}_j^T{\bf X}_j+k_2I\right)\widehat{\bm\Theta}_j(k_1)\rightarrow_p E(\Sigma) \mbox{ as } m\rightarrow\infty,$$ where $\Sigma$ is a random matrix. Let $F_{\min}$ be the distribution of the minimum eigenvector of the random matrix $\Sigma$. Thus, the approximate optimal choice is that $\eta_j,j=1,\cdots,N$, are independently generated from the distribution $F_\eta=F_{\min}$. This is only a theoretical argument, the implementation procedure is particularly complex. We suggest to use empirical methods as given above to choose the distribution $F_\eta$ because of the random projection theory and the robustness of the our method to the choice of $F_\eta$. For more details see the simulation study.
\end{enumerate}}

\subsection{The race-DC for general biased estimator}

In addition to that $\widehat{\bm\Theta}_j$ and $\widehat\sigma_j^{-2}$ can be empirically chosen, the initial estimator of $\beta$ can be chosen as a general biased estimator, not necessary the lasso estimator $\widehat\beta^{Lasso}_j$. The pivotal idea behind our technique is the use of the residual-adjustment structure in (\ref{(debiased-lasso)}) and a pro form linear regression model as in (\ref{(remodeling)}). Thus, the race-DC method developed in the previous subsection can be extended to the case of general biased estimation.

Let $\widehat\beta_j$ be a type of biased estimator of $\beta$ such as Lasso estimator or Ridge estimator or principal component estimator, computed on dataset ${\cal D}_j$. Similar to (\ref{(debiased-lasso)}), a residual-adjustment estimator is defined by \begin{eqnarray}\label{(debiased-general)}\widehat\beta^{RA}_j=\widehat\beta_j+\frac1m {\bf M}_j {\bf X}_j^T({\bf y}_j-{\bf X}_j\widehat\beta_j),\end{eqnarray} where ${\bf M}_j$ is a $p\times p$ matrix computed on dataset ${\cal D}_j$. The selection method for ${\bf M}_j$ is similar to that for $\widehat{\bm\Theta}_j$ given in the previous subsection.
Note that the residual-adjustment estimator (\ref{(debiased-general)}) can be recast as $$\widehat\beta^{RA}_j=\widehat\beta_j+\frac1m {\bf M}_j {\bf X}_j^T{\bf X}_j(\beta-\widehat\beta_j)+{\bm\epsilon}_j,$$
where ${\bm\epsilon}_j=\frac1m {\bf M}_j {\bf X}_j^T\bm\varepsilon_j$.
We then get the following pro form linear regression model:
\begin{eqnarray}\label{(remodeling1)}\widehat{z}_j={U}^T_j\beta+{\epsilon}_j,\ j=1,\cdots,N.\end{eqnarray} Here ${U}_j=\frac1m {\bf X}_j^T{\bf X}_j M^T_j\eta_j$ is $p$-dimensional covariate, $ \widehat{z}_j= \eta^T_i(\widehat\beta^{RA}_j-\widehat\beta_j)+{U}^T_j\widehat\beta_j$ is a scalar response variable, error ${\epsilon}_j=\eta^T_j\frac1m {\bf M}_j {\bf X}_j^T\bm\varepsilon_j$ and satisfies
\begin{eqnarray}\label{(error-2)}E[{\epsilon}_j|{\bf X}_j]=0\ \mbox{ and } \ Var[{\epsilon}_j|{\bf X}_j]=\frac{\sigma_\varepsilon^2}{m}\sigma^2_j\end{eqnarray} with $\sigma^2_j=\eta^T_j\frac{1}{m} {\bf M}_j {\bf X}_j^T{\bf X}_j{\bf M}_j^T\eta_j$.

Model (\ref{(remodeling1)}) has the same features as those in Remark 2.1.
For the model, the original {\it C2} is replaced with a general form as
\begin{enumerate}
\item[{\it C2'}.] The matrix $\sum_{j=1}^N\sigma_j^{-2}{U}_j{U}_j^T$ is positive definite, where \ $\sigma_j^{-2}$ is an approximate inverse to $\sigma^2_j$. \end{enumerate}
Similar to {\it C2}, here Condition {\it C2'} is easily satisfied as well.
Under Condition {\it C2'},
by solving the corresponding weighted least squares, we get the global estimator of $\beta$ as
\begin{eqnarray}\label{(GUE1)}\widetilde\beta^{GRA}=
\left(\sum_{j=1}^N \sigma_j^{-2} {U}_j{U}_j^T\right)^{-1}\sum_{j=1}^N
\sigma_j^{-2}{U}_j \widehat{z}_j.\end{eqnarray} Also this is a global residual-adjustment estimator (GRA estimator, for short). Similar to Remark 2.2, when $\frac1m{\bf X}_j^T{\bf X}_j$ for $j=1,\cdots,N$ are non-singular (or nearly non-singular) and ${\bf M}_j$ are chosen as $(\frac1m{\bf X}_j^T{\bf X}_j)^{-1}$ (or its Ridge form), $\widetilde\beta^{GRA}$ are exactly (or nearly) the ordinary least square estimator constructed by entire datasets. The following theorem states the consistency of $\widetilde\beta^{GRA}$, which guarantees the unbiasedness of $\widetilde\beta^{GRA}$ and the convergence rate, namely, the typical one $O(1/\sqrt{n})$.

\noindent{\bf Theorem 2.2.} {\it Under Conditions C1 and C2', the GRA estimator $\widetilde\beta^{GRA}$ in (\ref{(GUE1)}) satisfies
\begin{eqnarray}\label{(bias-variance)}E[\widetilde\beta^{GRA}|{\bf X}]=\beta \ \mbox{ and } \ Cov[\widetilde\beta^{GRA}|{\bf X}]=\frac{\sigma_\varepsilon^2}{n}\left(\frac1N\sum_{j=1}^N\sigma_j^{-2}{U}_j{U}_j^T\right)^{-1}.\end{eqnarray}}


\setcounter{equation}{0}
\section{The race-DC in nonlinear regression}\label{nonlinear}

In this section, we apply the race-DC algorithm to the nonlinear model. Suppose the data come from the following nonlinear model:
\begin{eqnarray}\label{(nonlinear)}Y_i=f(X_i,\beta) +\varepsilon_i, \ i=1,\cdots,n,\end{eqnarray} where $f(\cdot,\beta)$ is a known function and may be nonlinear in $\beta$, the other design conditions are the same as in linear model (\ref{(model)}). Let $\widehat\beta_j$ be a $\sqrt m$-consistent estimator of $\beta$ computed on ${\cal D}_j$, for example, the least squares estimator as
\begin{eqnarray}\label{(nonlinear-estimate)}\widehat\beta_j=\argmin_{\beta}\sum_{j\in{\cal H}_j}(Y_i-f(X_i,\beta))^2.\end{eqnarray}
It is known that the simple average of $\hat\beta_j$ and the DC expression by  AEE method of Lin and Xi
(2011) cannot achieve the $\sqrt{n}$-consistency under case of large $N$.
Motivated by the residual-adjustment estimators in (\ref{(debiased-lasso)}) and (\ref{(debiased-general)}), we define a residual-adjustment estimator for parameter $\beta$ in nonlinear regression as
\begin{eqnarray}\label{(nonlinear-estimate1)}\widehat\beta^{RA}_j=\widehat\beta_j+\frac1mH_j({\bf y}_j-f({\bf X}_j,\widehat\beta_j)),\end{eqnarray}
where $H_j$ is a given $p\times m$ matrix depending only on dataset ${\cal D}_j$, and $f({\bf X}_j,\beta)=(f(X_i,\beta):i\in{\cal H}_j)^T$. Similar to the case of linear model, $H_j$ can be empirically determined, and particularly, it can be chosen as $\widehat {\bf M}_j\dot f^T({\bf X}_j,\widehat\beta_j)$, where  $\dot f({\bf X}_j,\beta)=(\dot f^T(X_i,\beta):i\in{\cal H}_j)^T$ is $m\times p$ matrix with $\dot f(x,\beta)$ being the derivative of $f(x,\beta)$ with respect to $\beta$, and $\widehat {\bf M}_j$ is an approximate reverse matrix to $\dot f^T({\bf X}_j,\widehat\beta_j)\dot f({\bf X}_j,\widehat\beta_j)$.
Then, similar to the pro form regressions (\ref{(remodeling)}) and (\ref{(remodeling1)}), the following regression model holds:
\begin{eqnarray}\label{(nonlinear-remodeling)}\widehat{z}_j=\eta^T_j\frac1mH_jf({\bf X}_j,\beta)+{\epsilon}_j,\ j=1,\cdots,N.\end{eqnarray} where $ \widehat{z}_j=\eta^T_j( \widehat\beta^{RA}_j-\widehat\beta_j+\frac1mH_jf({\bf X}_j,\widehat\beta_j))$ is a scalar response variable and  ${\epsilon}_j=\frac1m\eta^T_jH_j\bm\varepsilon_j$ is a new error term satisfying
\begin{eqnarray}\label{(nonlinear-error)}E[{\epsilon}_j|{\bf X}_j]=0\ \mbox{ and } \ Var[{\epsilon}_j|{\bf X}_j]=\frac{\sigma_\varepsilon^2}{ m}\sigma^2_j \end{eqnarray} with $\sigma^2_j=\frac1m\eta^T_jH_jH_j^T\eta_j.$ According to the nonlinear regression (\ref{(nonlinear-remodeling)}) together with expectation and variance condition (\ref{(nonlinear-error)}),
the corresponding least squares estimator is defined by
\begin{eqnarray}\label{(nonlinear-LS)}\widetilde\beta^{GRA}=\argmin_{\beta}\sum_{j=1}^N
\sigma^{-2}_j\left(\widehat{z}_j-
\eta^T_j\frac1mH_jf({\bf X}_j,\beta)\right)^2,\end{eqnarray} where $\sigma^{-2}_j$ is an approximate reverse to $\sigma^2_j$. Also this is a global residual-adjustment estimator (GRA estimator, for short).
To get a valid least squares solution, we need the following condition: \begin{enumerate}
\item[{\it C2"}.] The matrix $\sum_{j=1}^N\sigma^{-2}_jW_j(\beta)W^T_j(\beta)$ is positive definite in a neighbourhood of the true value of $\beta$, where $W_j(\beta)=\frac1m\dot f^T({\bf X}_j,\beta)H^T_j\eta_j$, a $p$-dimensional column vector.\end{enumerate}
Similar to Conditions {\it C2} and {\it C2'}, here Condition {\it C2"} is easily be satisfied as well.

Unlike the case of linear model, the least squares estimator (\ref{(nonlinear-LS)}) in nonlinear regression (\ref{(nonlinear)}) has no closed-form expression. We suggest the following algorithm to compute the numerical solution of
the GRA estimator $\widetilde\beta^{GRA}$:
\begin{enumerate}
\item[] {\it Step 1.}
Choose an initial value $\beta^{(0)}$ of $\beta$, for example $\beta^{(0)}=\widehat\beta_1$, the least squares estimator computed on ${\cal D}_1$.

{\it Step 2.} Each computer calculates the corresponding compressed data $H_j{\bf y}_j$ and $H_jf({\bf X}_j,\beta^{(0)})$, and sends them to the central computer.

{\it Step 3.} Based on the initial choice $\beta^{(0)}$ of $\beta$, and the compressed data from each computer, the central computer implements random projection and calculates the first iterative solution $\beta^{(1)}$ by least squares (\ref{(nonlinear-LS)}) together with its Newton-Raphson algorithm, and then sends $\beta^{(1)}$ back to each computer.

{\it Step 4.} Use $\beta^{(1)}$ as an updated initial value of $\beta$, and repeat {\it step 1} - {\it step 3} until a stop rule is met. Then, the last iterative solution is the final numerical result of the estimator.\end{enumerate}

It will be seen in simulation study that with rather few iterations (generally, the number of iterations is 3; see simulation study), we can get a satisfactory numerical result if the regression function $f(x,\beta)$ is smooth. On the other hand, regarding the property of DC-expression, and $\sqrt n$-consistency of the GRA estimator $\widetilde\beta^{GRA}$ given in (\ref{(nonlinear-LS)}), we have the following conclusions:

{\bf Remark 3.1.} {\it
\begin{enumerate}
\item[(1)] DC-expression of numerical solution. As was shown above that the solution $\widetilde\beta^{GRA}$ to the optimization problem in (\ref{(nonlinear-LS)}) has no closed-form expression, but can be easily obtained by common numerical methods.
The objective function in (\ref{(nonlinear-LS)}) is a DC-expression, as a result the iteration algorithm is also a DC-expression, depending only on the compressed data.
\item[(2)] $\sqrt n$-consistency.
Model (\ref{(nonlinear-remodeling)}) is a common nonlinear regression. The main difference from the classical one is that the variances of errors ${\epsilon}_j$ in (\ref{(nonlinear-error)}) are of order $O(1/m)$, rather than a constant. Then, the resulting global estimator is $\sqrt n$-consistent, i.e.,
\begin{eqnarray}\label{(consistency)}\widetilde\beta^{GRA}-\beta=O_p(1/\sqrt n).\end{eqnarray} A brief proof for (\ref{(consistency)}) is given as follows.
Note that under some regularity conditions (e.g., the smoothness of $f(x,\beta)$), the solution $\widetilde\beta^{GRA}$ to (\ref{(nonlinear-LS)}) satisfies the estimating equation:
$$\frac1N\sum_{j=1}^N
\sigma^{-2}_j W_j(\widetilde\beta^{GRA})\left(\widehat{z}_j-
\eta^T_j\frac1mH_jf({\bf X}_j,\widetilde\beta^{GRA})\right)=0,$$ where $W_j(\beta)$ is the derivative of $\eta^T_j\frac1mH_jf({\bf X}_j,\beta)$ with respect to $\beta$ as defined in {\it C2"}. Under model (\ref{(nonlinear-remodeling)}), the above can be rewritten as
$$\frac1N\sum_{j=1}^N
\sigma^{-2}_j W_j(\widetilde\beta^{GRA})\left(\eta^T_j\frac1mH_jf({\bf X}_j,\beta)+{\epsilon}_j-
\eta^T_j\frac1mH_jf({\bf X}_j,\widetilde\beta^{GRA})\right)=0.$$ According to the property of least squares estimator in nonlinear regression (see, e.g., Seber and Wild; 2003), the least squares estimator $\widetilde\beta^{GRA}$ is consistent under some regularity conditions (for example the continuity of the second-derivative of $f(x,\beta)$). Then, by Taylor expansion and the property $\epsilon_j=O_p(1/\sqrt m)$, we have
\begin{eqnarray*}&&\frac1N\sum_{j=1}^N
\sigma^{-2}_j W_j(\beta)\left(W^T_j(\beta)(\widetilde\beta^{GRA}-\beta)-{\epsilon}_j
+o_p(\|\widetilde\beta^{GRA}-\beta\|_2)\right)\\&&+o_p(\|\widetilde\beta^{GRA}-\beta\|_2)=0.\end{eqnarray*} Consequently.
\begin{eqnarray*}\widetilde\beta^{GRA}-\beta&=&\left(\frac1N\sum_{j=1}^N\sigma^{-2}_j W_j(\beta)W^T_j(\beta)\right)^{-1}\frac1N\sum_{j=1}^N\sigma^{-2}_j W_j(\beta){\epsilon}_j\\&&+o_p(\|\widetilde\beta^{GRA}-\beta\|_2),\end{eqnarray*}
As mentioned above, for given $\bf X$, ${\epsilon}_j$ have conditional expectation zero and conditional variance variances of order $O(1/m)$. Thus, the dominating
term on the right hand of the above formula has conditional expectation zero and conditional covariance of order $O(1/n)$, which implies (\ref{(consistency)}). \end{enumerate}

 }

\section{Simulation studies}
In this section, a series of simulations are conducted to comprehensively evaluate the performance of the proposed race-DC method. To this end, the following biased estimators are considered as basic objects for bias-correction: Lasso estimator, Ridge estimator and principal component estimator (PCE) in linear regression, and least squares estimator in nonlinear regression. The newly proposed global estimators are compared to various competitors such as naive average of residual adjustment estimator, the DC-expressions of different estimators (see the definitions given below). The estimator based on entire dataset is used as a benchmark for comparison. Various model settings under identically distributed and non-identically distributed data are considered in the experiment procedure.
The estimation bias and the mean squared error (MSE) are employed to evaluate the performances of all the methods, and each experiment is repeated 500 times.

\subsection{Linear model}
{\it Experiment 1. Lasso-based estimators.} We first check the performance of race-DC using Lasso estimator. Let the data be generated from the following linear model:
\begin{equation}
Y_i = X_i^T\beta + \varepsilon_i,\ i=1,\cdots,n,
\end{equation}
where $\beta=(3,2,1,0.5,-2,0,\cdots,0)$ is a 30-dimensional vector, $\varepsilon_i,i=1,\cdots,n$, are independent and identically distributed as $N(0,4)$,  the predictor is generated respectively from the following two settings:
\begin{enumerate}
	\item [(a)]Identically distributed case: $X_1,\cdots,X_n$, are independent, and $X_i\in\mathcal{D}_j$ for all $j$ comes from the same distribution $N(0,\Sigma)$ with the $(i,j)$-th element of $\Sigma$ equal to $0.5^{|i-j|}$;
	\item [(b)]Non-identically distributed case: $X_1,\cdots,X_n$, are independent, and $X_i\in\mathcal{D}_j$ for given $j$ comes from $ N(\mu_j,\Sigma)$, where $\mu_j,j=1,\cdots,N$, are randomly generated from $N(0,I_p)$ and $\Sigma$ keeps the same setting as in case (a).
\end{enumerate}

This is a sparse linear model with relatively high-dimensional predictor. It shows that we can use Lasso-based methods to select variables and estimate parameters, simultaneously.
Thus, we compare the race-DC with two competitors: the simple average of residual adjustment Lasso estimator $\overline\beta^{GDBL}$ as in (\ref{(average)}),  and the DC expression of Lasso estimator defined by
\begin{equation}\label{dclasso}
\widehat\beta^{Lasso} = \left(\sum_{j=1}^N\frac1m{\bf X}_j^T{\bf X}_j\right)^{-1}\sum_{j=1}^N\frac1m{\bf X}_j^T{\bf X}_j\widehat\beta_j^{Lasso},
\end{equation}
where $\widehat\beta_j^{Lasso}$ is the lasso solution to (\ref{(lasso)}). The regularization parameters $\lambda_j$ for $j=1,\cdots,N$, are selected by 5-fold cross-validation. For the details see Meinshausen and Buhlmann (2010), Shah and Samworth (2013), and Chen and Xie (2014). We set the sample size $n$ equal to 10000 and the number of batches $N$ equals to 50, 100, 200 and 400, respectively. Note that the ordinary least square will not work any more within each data batch when $N=400$.

When implementing the race-DC, we use the method of random sampling to generate $\eta_j$, i.e., $\eta_j$ is generated from a certain distribution. Particularly, in our simulation, each component in $\eta_j$ is sampled from the normal distribution $N(0,1/p)$, where $p$ is the dimension of the predictor $X$.
We perform this kind of sampling 200 times and set the parameter estimation equal to the average of the results got from each time of projection. Note that the random projection algorithm is implemented in the central machine and can be calculated rapidly. Besides, throughout this simulation, we set $k_1=0.1$.

\begin{figure}[H]
	\centering
	\includegraphics[width=6in]{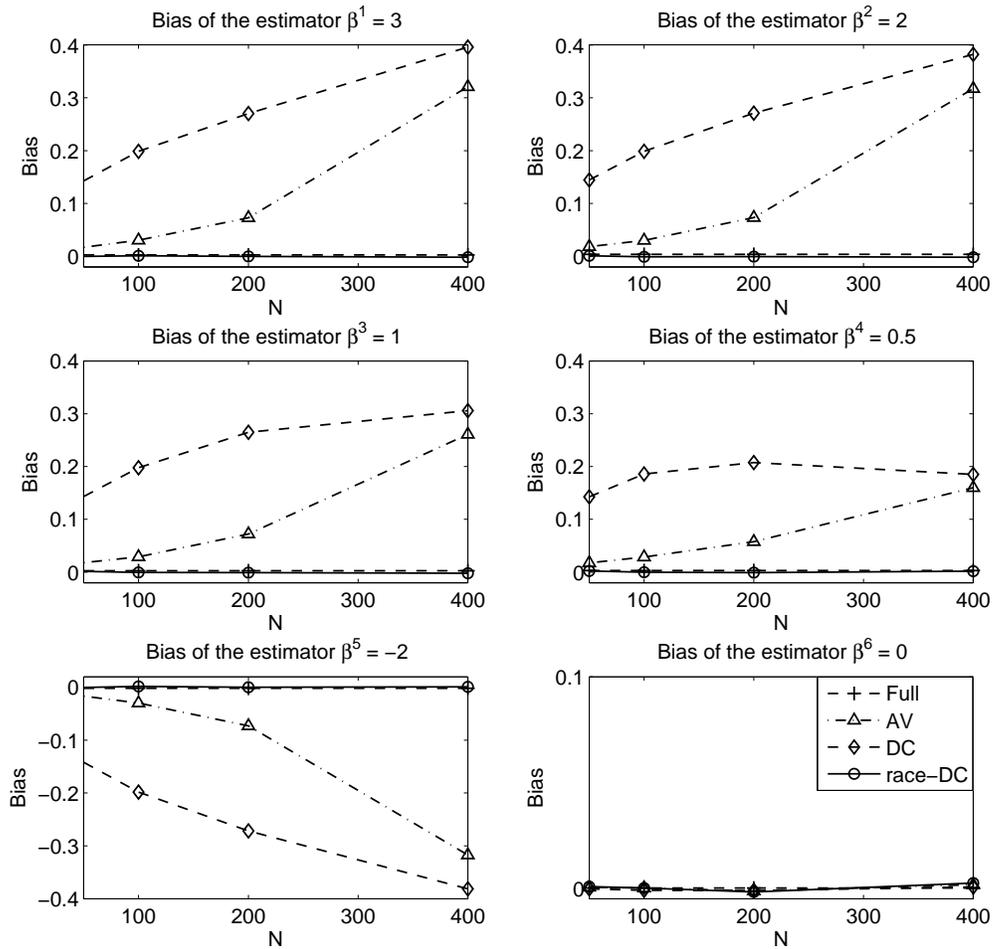}\vspace{-6ex}\\
	\caption{Bias of Lasso-based estimators for identically distributed data in Experiment 1}\label{biaslassoiid}
\end{figure}

\begin{figure}[H]
	\centering
	\includegraphics[width=6in]{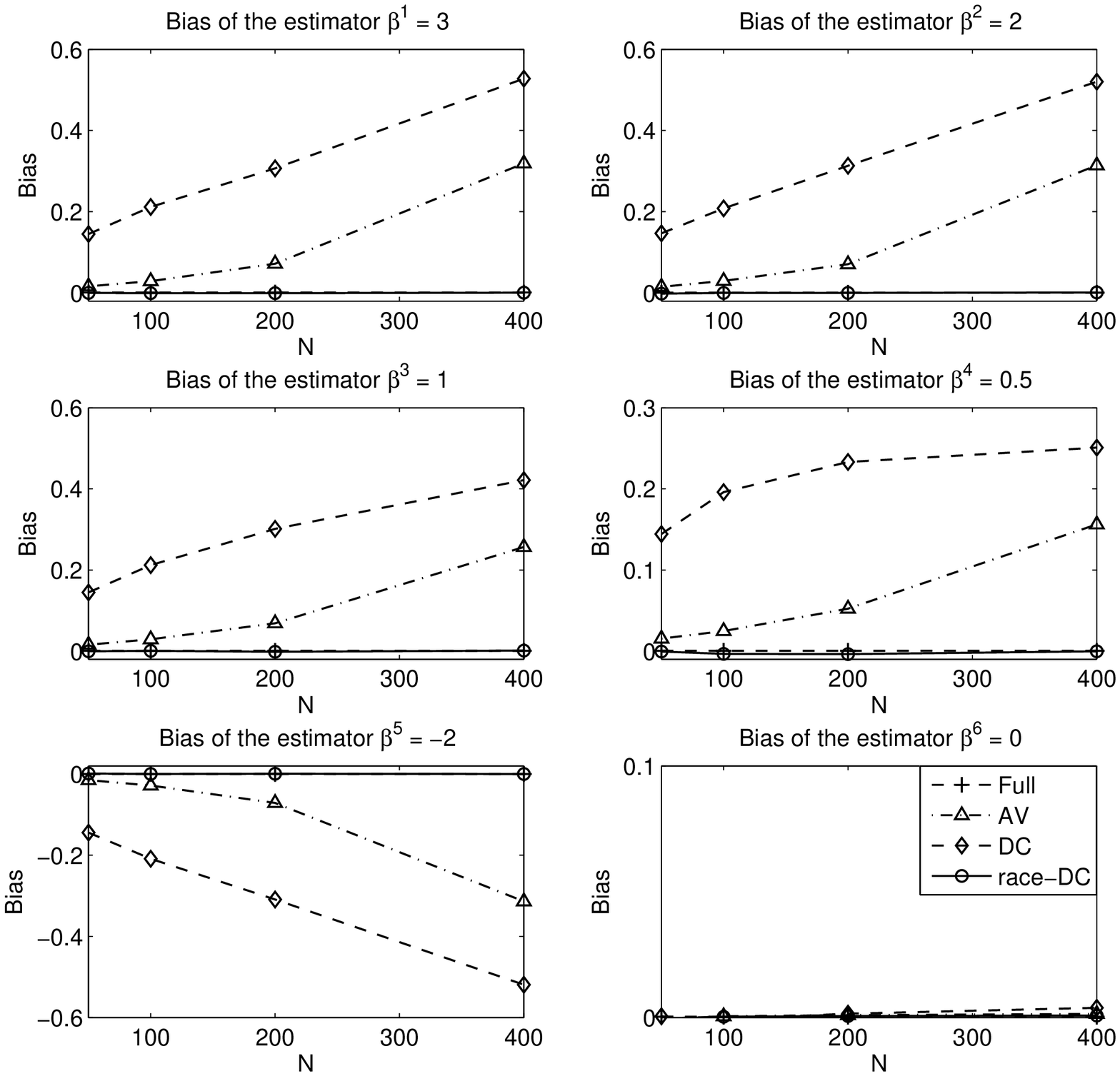}\vspace{-6ex}\\
	\caption{Bias of Lasso-based estimators for non-identically distributed data in Experiment 1}\label{biaslassononiid}
\end{figure}

\begin{figure}[H]
	\centering
	\includegraphics[width=6in]{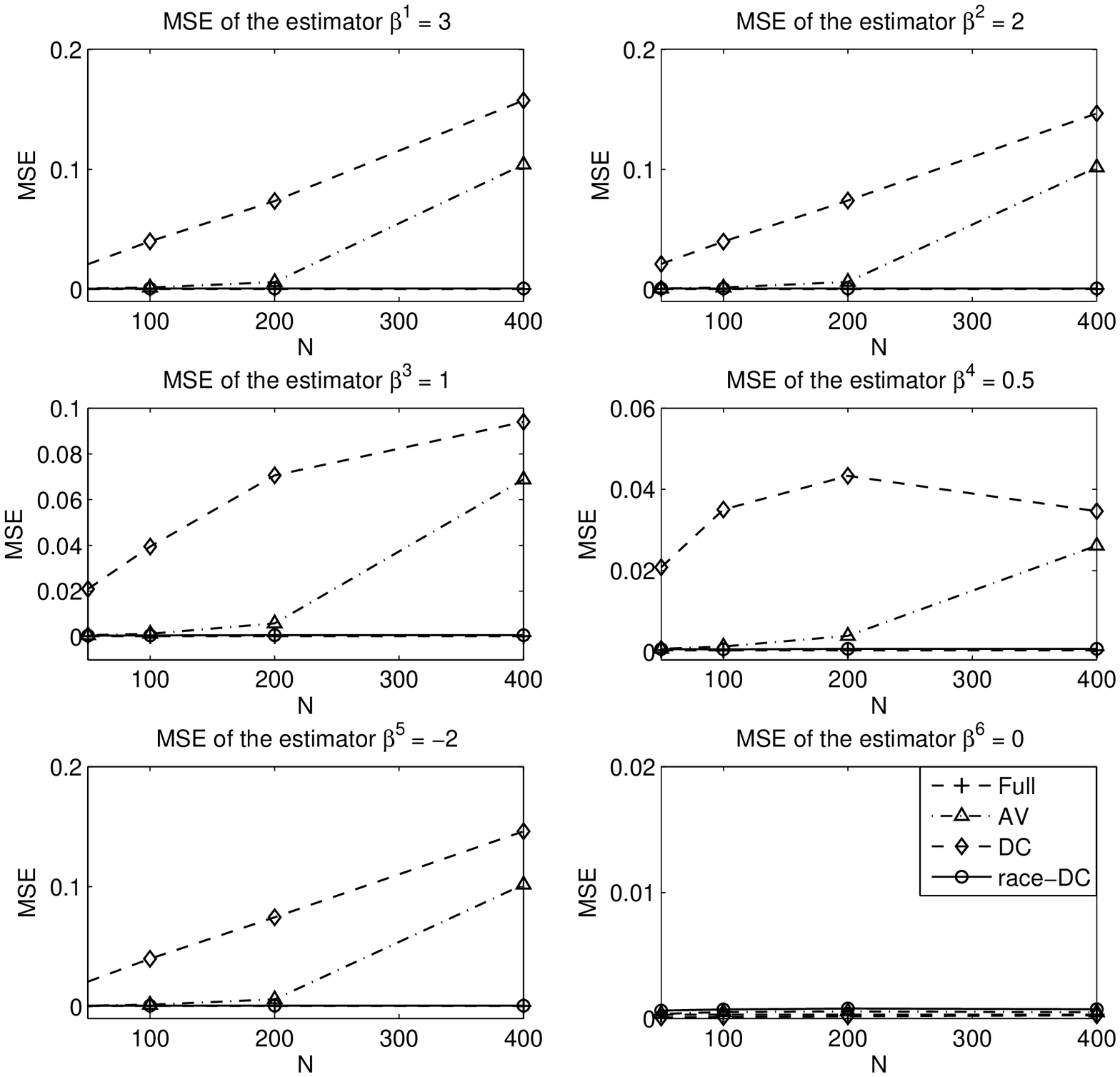}\vspace{-6ex}\\
	\caption{MSE of Lasso-based estimators for identically distributed data in Experiment 1}\label{mselassoiid}
\end{figure}

\begin{figure}[H]
	\centering
	\includegraphics[width=6in]{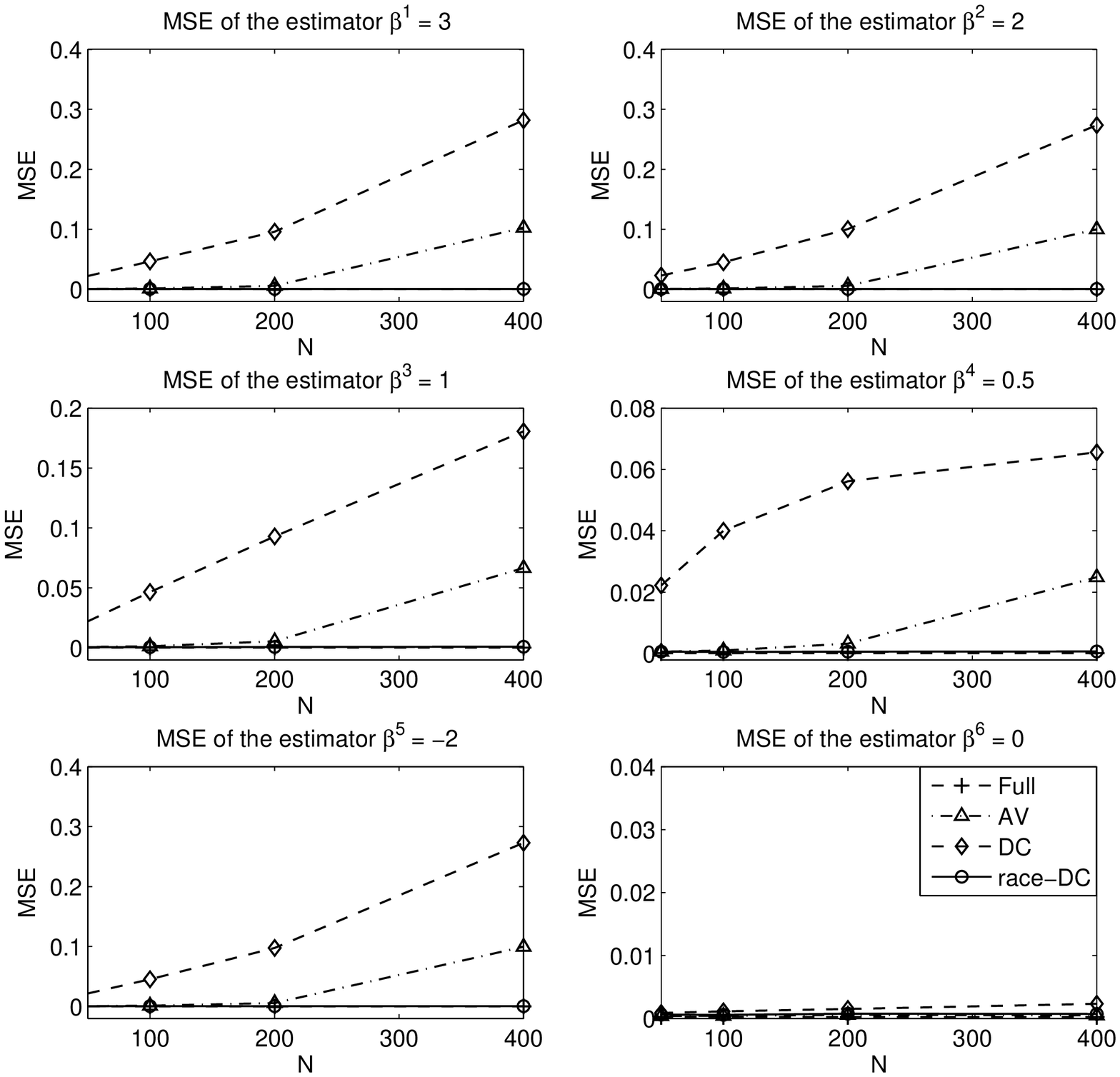}\vspace{-6ex}\\
	\caption{MSE of Lasso-based estimators for non-identically distributed data in Experiment 1}\label{mselassononiid}
\end{figure}

Figures \ref{biaslassoiid}-\ref{biaslassononiid} display the bias of the resultant estimators and Figures \ref{mselassoiid}-\ref{mselassononiid} display the corresponding MSE of the estimators, where the legend AV represents the simple average of residual adjustment Lasso estimator $\overline\beta^{GDBL}$, DC represents the estimator defined in (\ref{dclasso}) and the Full represents the residual adjustment Lasso estimator constructed by entire dataset.
\begin{center}
Figures \ref{biaslassoiid}-\ref{mselassononiid} about here (attached at the end of the manuscript)
\end{center}
From these figures, the following results can be easily observed:
\begin{enumerate}
	\item [(1)] Under all the model settings,
the race-DC performs very well with a very small bias and MSE, and is significantly better than the AV and the DC method for any choice of $N$. Moreover, the difference between the race-DC and the Full method is negligible.
\item [(2)] The performance of AV method becomes worse when $N$ grows, which is consistent with the theoretical conclusion that the simple average only works under small $N$, but would incur a non-negligible bias under the case of large $N$.
\item [(3)] The DC method behaves worse than the AV. It is because the estimation bias of the DC method is not corrected during the composition procedure, no matter locally or globally.
\item [(4)] The non-identically distributed data have a negative effect on the two competitors but hardly does not have any effect on the race-DC method. Specifically, when the data are not identically distributed, the bias and MSE of the two competitors will be enlarged while those of the race-DC still keeps around zero.
\item [(5)] All the methods estimate the zero coefficients equal to zero, approximately. More specifically, by taking the hard or soft thresholdings as mentioned in (\ref{(threshold)}), the estimators of zero coefficients can be exactly shrunk to zero by setting $t=\sqrt{\log p/n}$.
\end{enumerate}

\

\noindent
{\it Experiment 2. Ridge-based estimators.} We now take the Ridge estimator as a basic object and use the following model to check the effectiveness of the race-DC:
\begin{equation}
Y_i = X_i^T\beta + \varepsilon_i, \ i=1,\cdots,n,
\end{equation}
where $\beta=(2,1.5,1,0.5,-2,0)$, $X_1,\cdots,X_n$, are independent, and $X_i\in\mathcal{D}_j$ for all $j$ comes from the same distribution $N(0,\Sigma)$ with the $(i,j)$-th element equal to $0.95^{I(i\neq j)}$, and $\varepsilon_i,i=1,\cdots,n$, are independent and identically distributed as $N(0,4)$. Here and in the following, we only show the results of identically distributed preditors, one can refer to Supplementary materials for the results of non-identically distributed data.

\begin{figure}[H]
	\centering
	\includegraphics[width=6in]{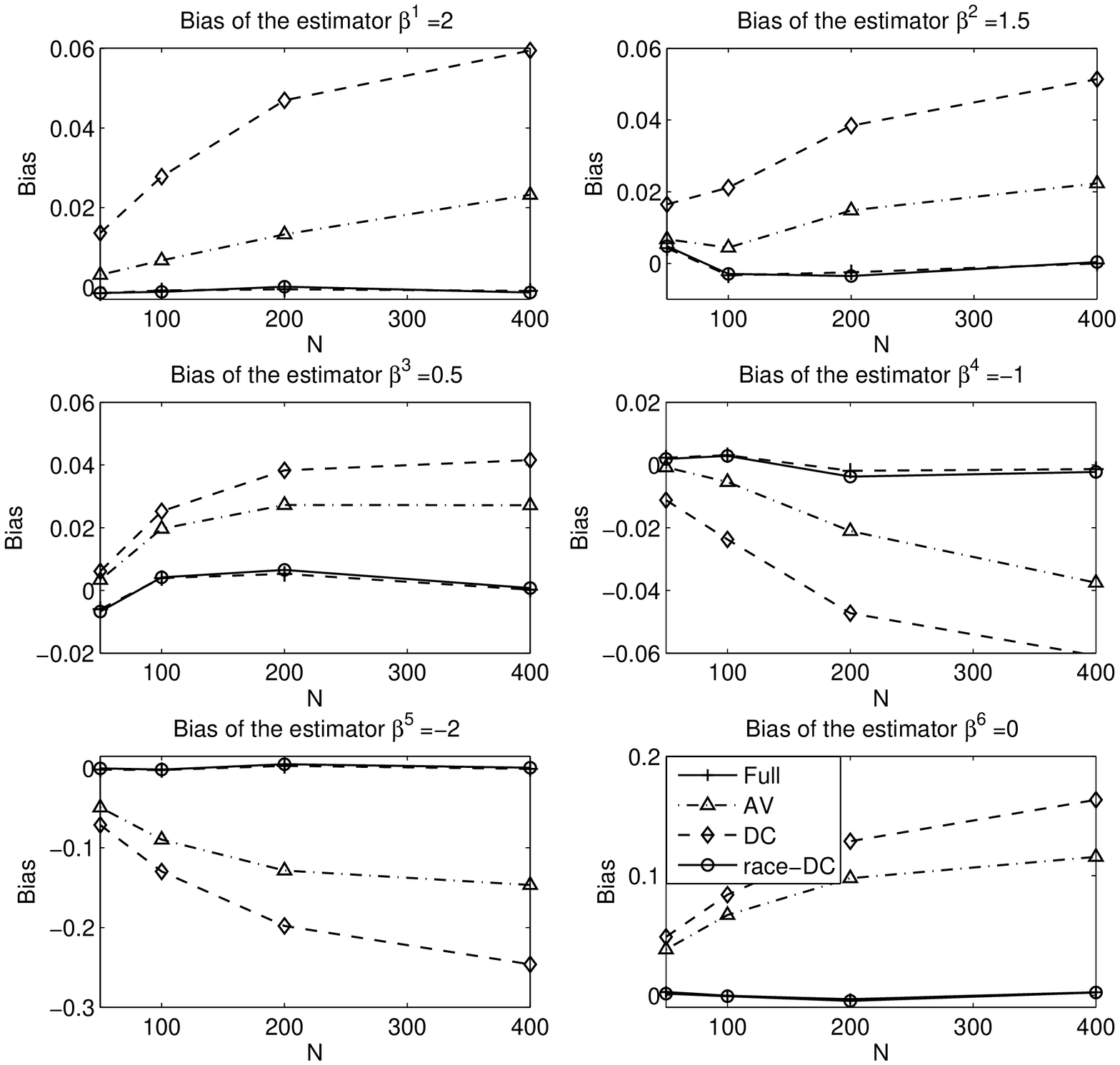}\vspace{-6ex}\\
	\caption{Bias of Ridge-based estimators with the regularization parameter determined by HK method in Experiment 2}\label{biasridgeHK}
\end{figure}

\begin{figure}[H]
	\centering
	\includegraphics[width=6in]{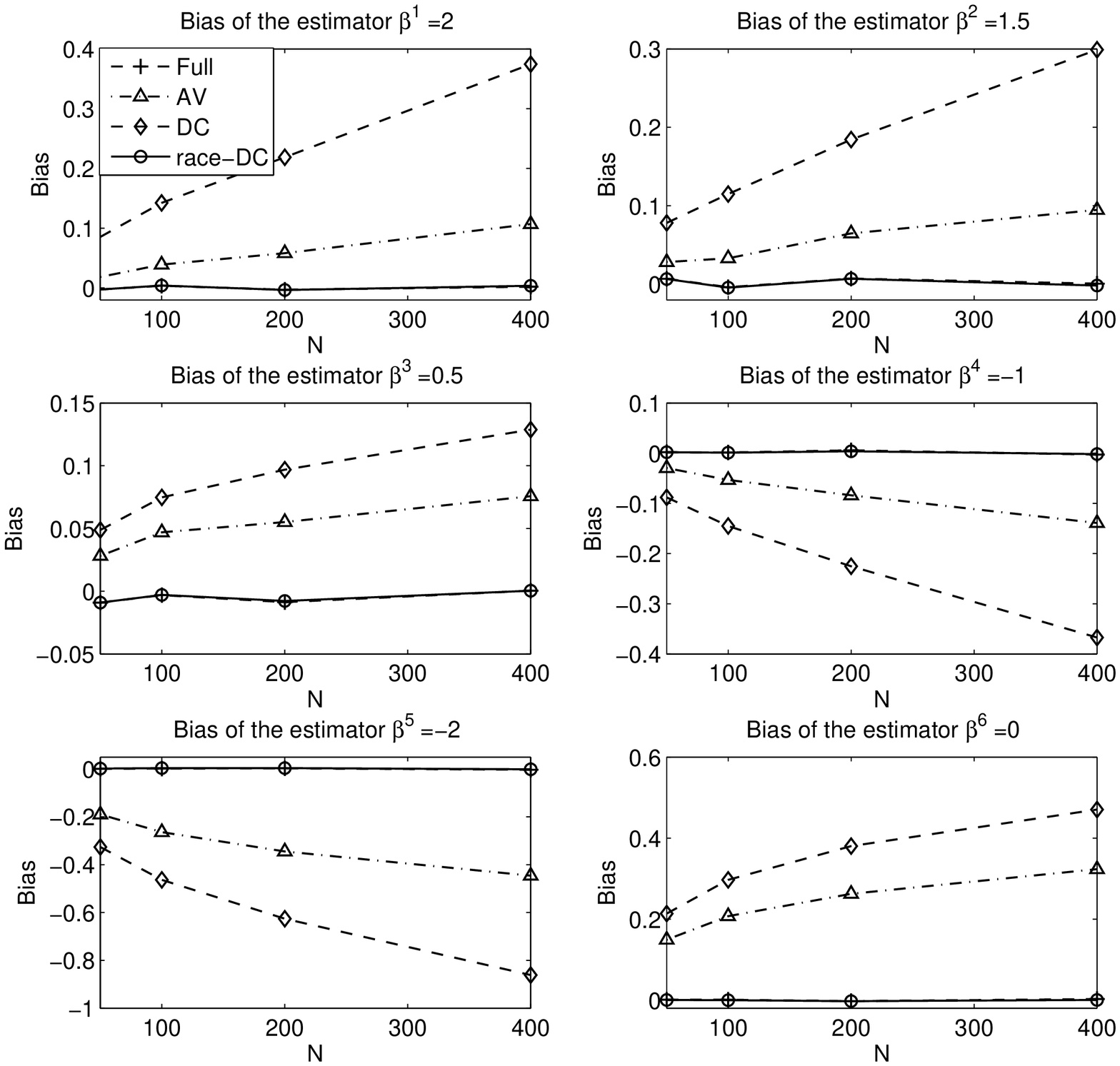}\vspace{-6ex}\\
	\caption{Bias of Ridge-based estimators with the regularization parameter determined by CV criterion in Experiment 2}\label{biasridgeCV}
\end{figure}

\begin{figure}[H]
	\centering
	\includegraphics[width=6in]{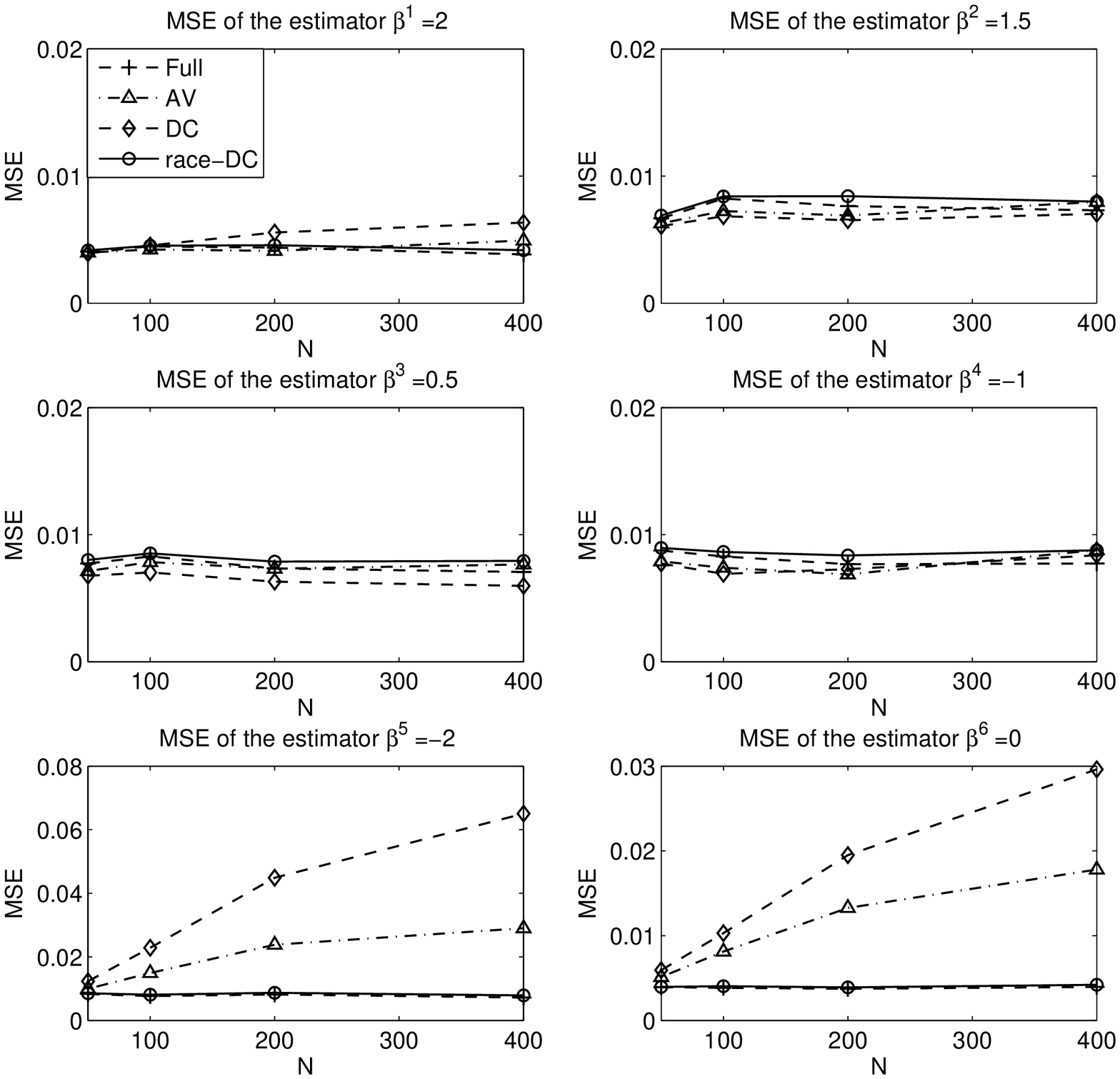}\vspace{-6ex}\\
	\caption{MSE of Ridge-based estimators with the regularization parameter determined by HK method in Experiment 2}\label{mseridgeHK}
\end{figure}

\begin{figure}[H]
	\centering
	\includegraphics[width=6in]{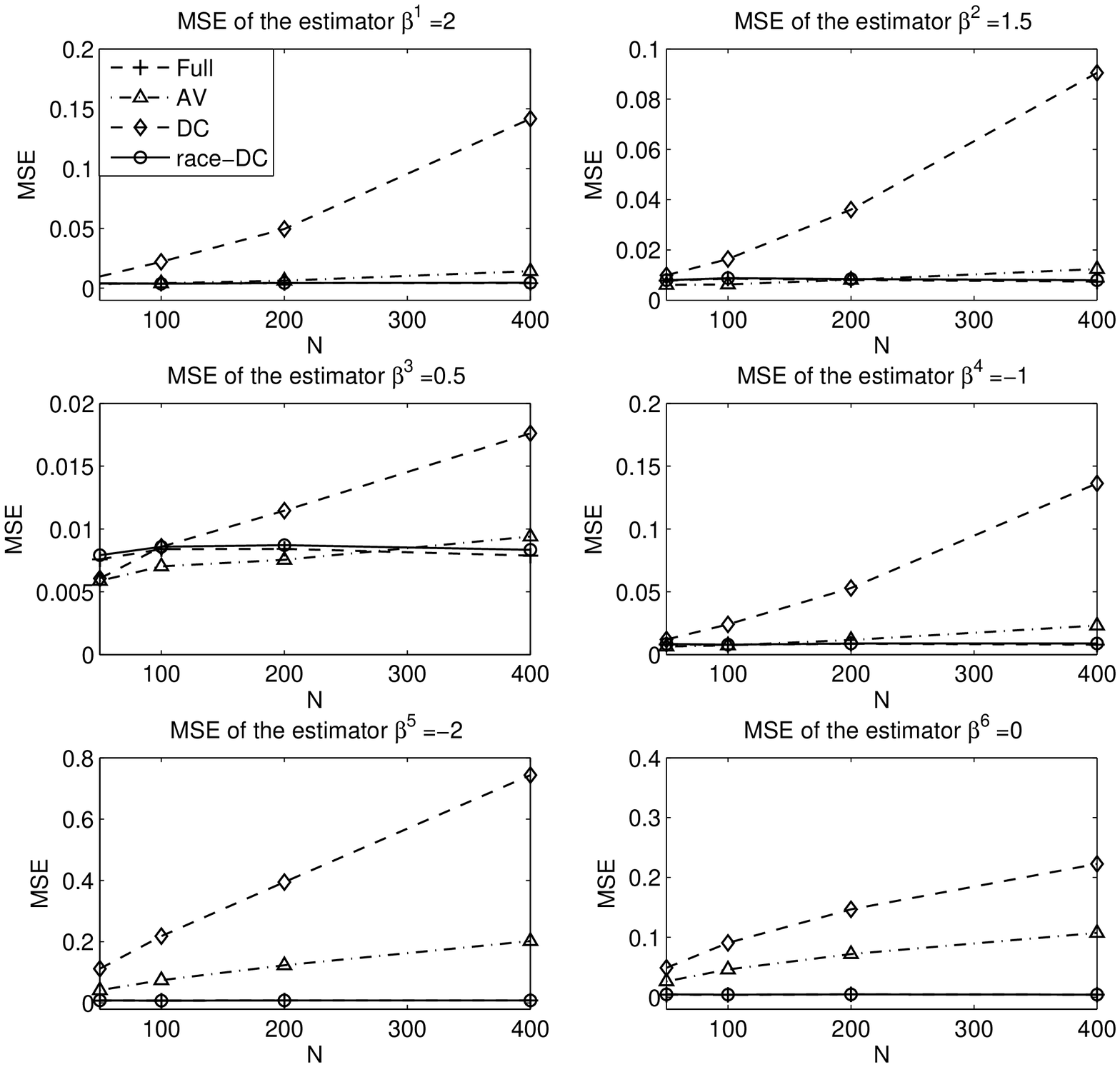}\vspace{-6ex}\\
	\caption{MSE of Ridge-based estimators with the regularization parameter determined by CV criterion in Experiment 2}\label{mseridgeCV}
\end{figure}

In this linear model, the dimension of parameter is low and the correlation between the predictors is very high. This shows that we can use Ridge-based methods to estimate parameters.
Similar to (\ref{(average)}), we compare our race-DC with two competitors: the simple average of residual-adjustment Ridge estimator defined by
\begin{equation*}
\overline\beta^{GDBR}=\frac1N\sum_{j=1}^N\widehat\beta_j^{DBR},
\end{equation*}
where $
\widehat\beta^{DBR}_j=\widehat\beta_j^{Ridge}+\frac1m {\bf M}_j {\bf X}_j^T({\bf y}_j-{\bf X}_j\widehat\beta_j^{Ridge})
$
with ${\bf M}_j=\left(\frac1m{\bf X}_j^T{\bf X}_j+k_1I\right)^{-1}$, and the DC expression of Ridge estimator (See Lin and Xi, 2011) defined as
\begin{equation*}
\widehat\beta^{Ridge} = \left(\sum_{j=1}^N{\bf A}_j\right)^{-1}\sum_{j=1}^N{\bf A}_j\widehat\beta_j^{Ridge},
\end{equation*}
where ${\bf A}_j={\bf X}_j^T{\bf X}_j+s_jI$, and $\widehat\beta_j^{Ridge}$ is the Ridge estimator computed on ${\cal D}_j$ as $({\bf X}_j^T{\bf X}_j+s_jI)^{-1}{\bf X}_j{\bf y}$ with ridge value $s_j>0$. Here the regularization parameter $s_j$ is specified by users. Particularly, in our simulation, we determine $s_j$ by two different ways, one is the Horel-Kennard (HK) formula (Horel and Kennard, 1970), the other is the $K$-fold cross-validation (CV) criterion.  We set $K=5$ in our simulation.
\begin{center}
	Figures \ref{biasridgeHK}-\ref{mseridgeCV} about here (attached at the end of the manuscript)
\end{center}
The simulations are shown in Figures \ref{biasridgeHK}-\ref{mseridgeCV}.
From these figures, some similar phenomena to those in (1)-(5) of Experiment 1 can be observed, the details are omitted here. Besides, we have the following different findings. Note that the HK method tends to select a small ridge value while CV method selects a relative large ridge value. The ridge value has a non-ignorable impact on the bias and MSE. Generally, the estimation bias of the CV criterion is significantly larger than that of HK method.
On the other hand, the race-DC and the Full method can relatively precisely estimate the zero coefficient $\beta_6$, but the estimators of the zero coefficient by the two competitors have a large bias, specifically for the case of large $N$.

{\it Experiment 3. Principle component-based estimators (PCE).} The goal of this experiment is to examine the performance of the race-DC using PCE as a basic estimator. The data are generated from the following model:
\begin{equation}
Y_i = X_i^T\beta + \varepsilon_i, \ i=1,\cdots,n,
\end{equation}
where $\beta=(3,2,1,-1,-2,0)$, $X_1,\cdots,X_n$, are independent and identically distributed as $N(0,\Sigma)$ with the $(i,j)$-th element for $i\neq j$ equal to $0.9^{I(i\neq j)}$, and $\varepsilon_1,\cdots,\varepsilon_n$, are independent of identically distributed as $N(0,0.25)$.

\begin{figure}[H]
	\centering
	\includegraphics[width=6in]{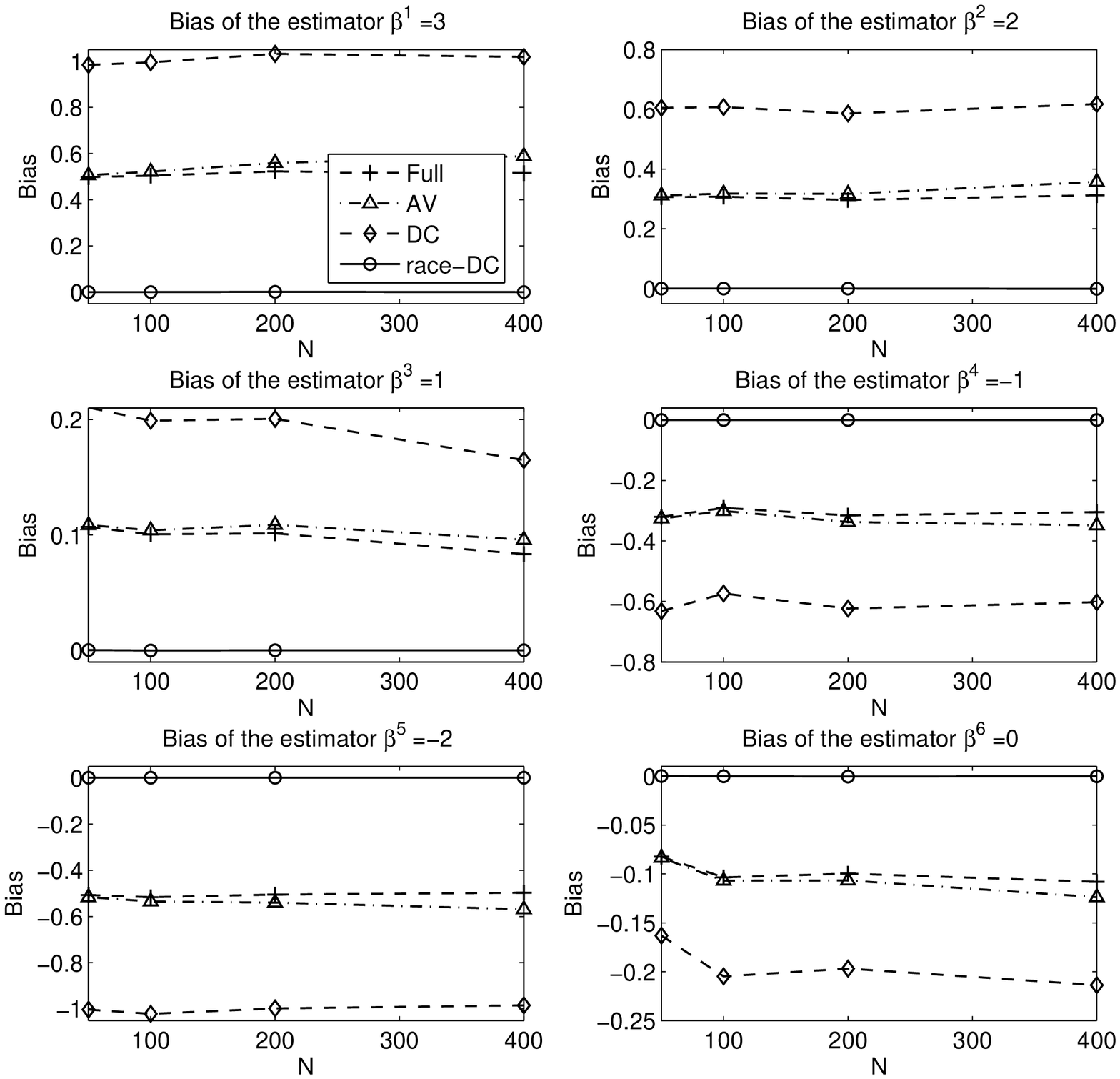}\vspace{-6ex}\\
	\caption{Bias of PCE with $r=4$ in Experiment 3}\label{biaspcet4}
\end{figure}

\begin{figure}[H]
	\centering
	\includegraphics[width=6in]{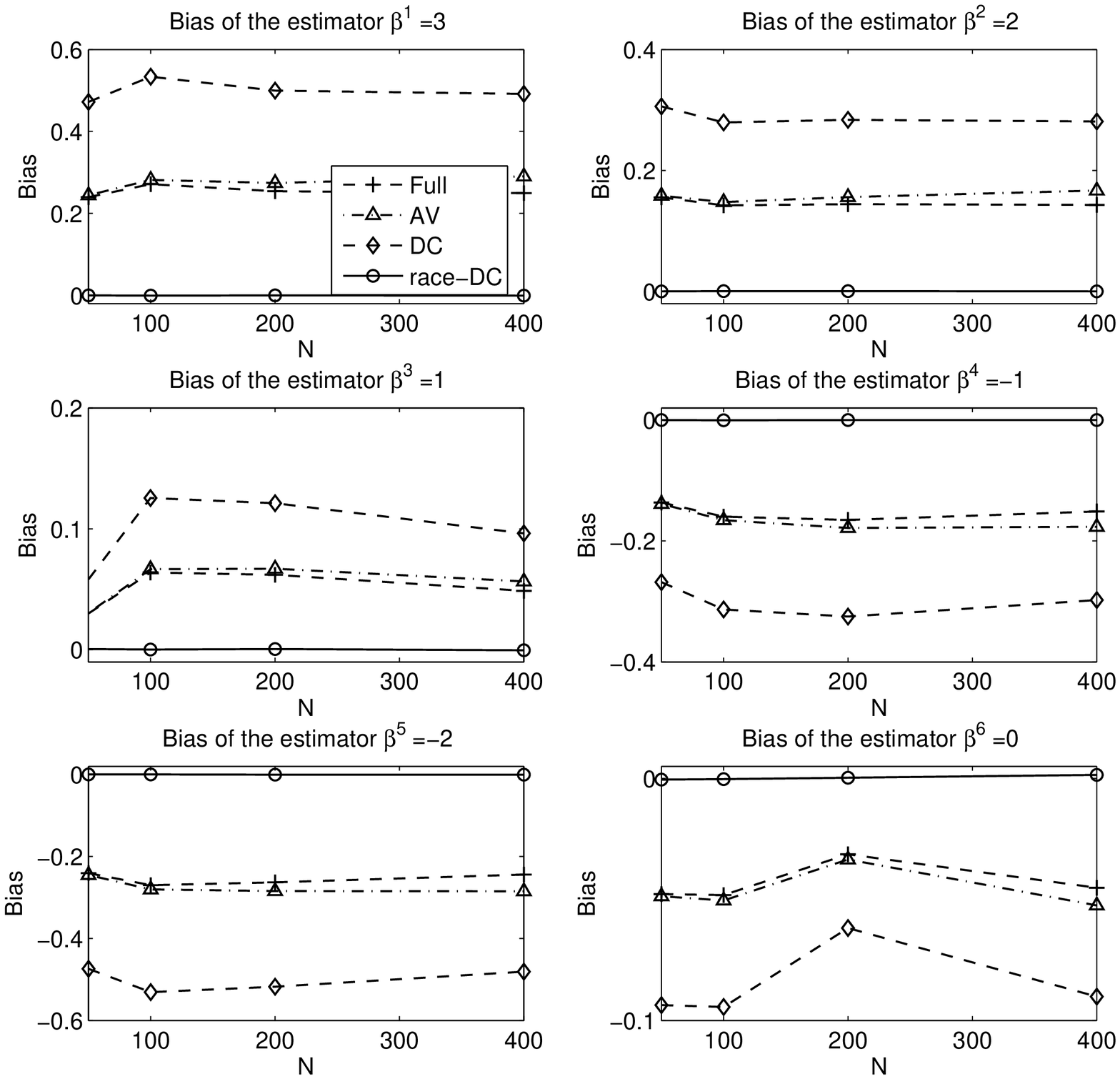}\vspace{-6ex}\\
	\caption{Bias of PCE with $r=5$ in Experiment 3}\label{biaspcet5}
\end{figure}

\begin{figure}[H]
	\centering
	\includegraphics[width=6in]{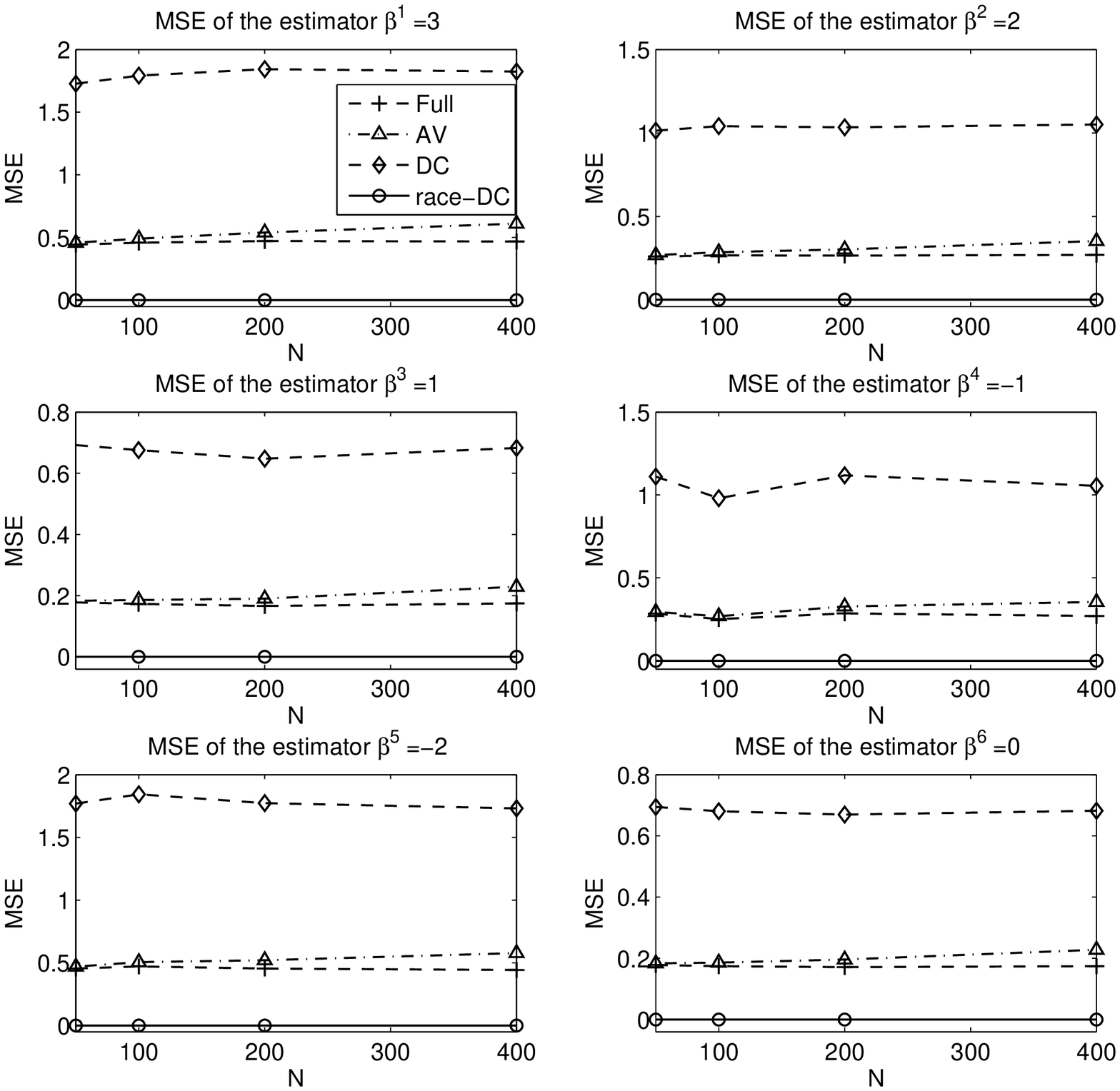}\vspace{-6ex}\\
	\caption{MSE of PCE with $r=4$ in Experiment 3}\label{msepcet4}
\end{figure}

\begin{figure}[H]
	\centering
	\includegraphics[width=6in]{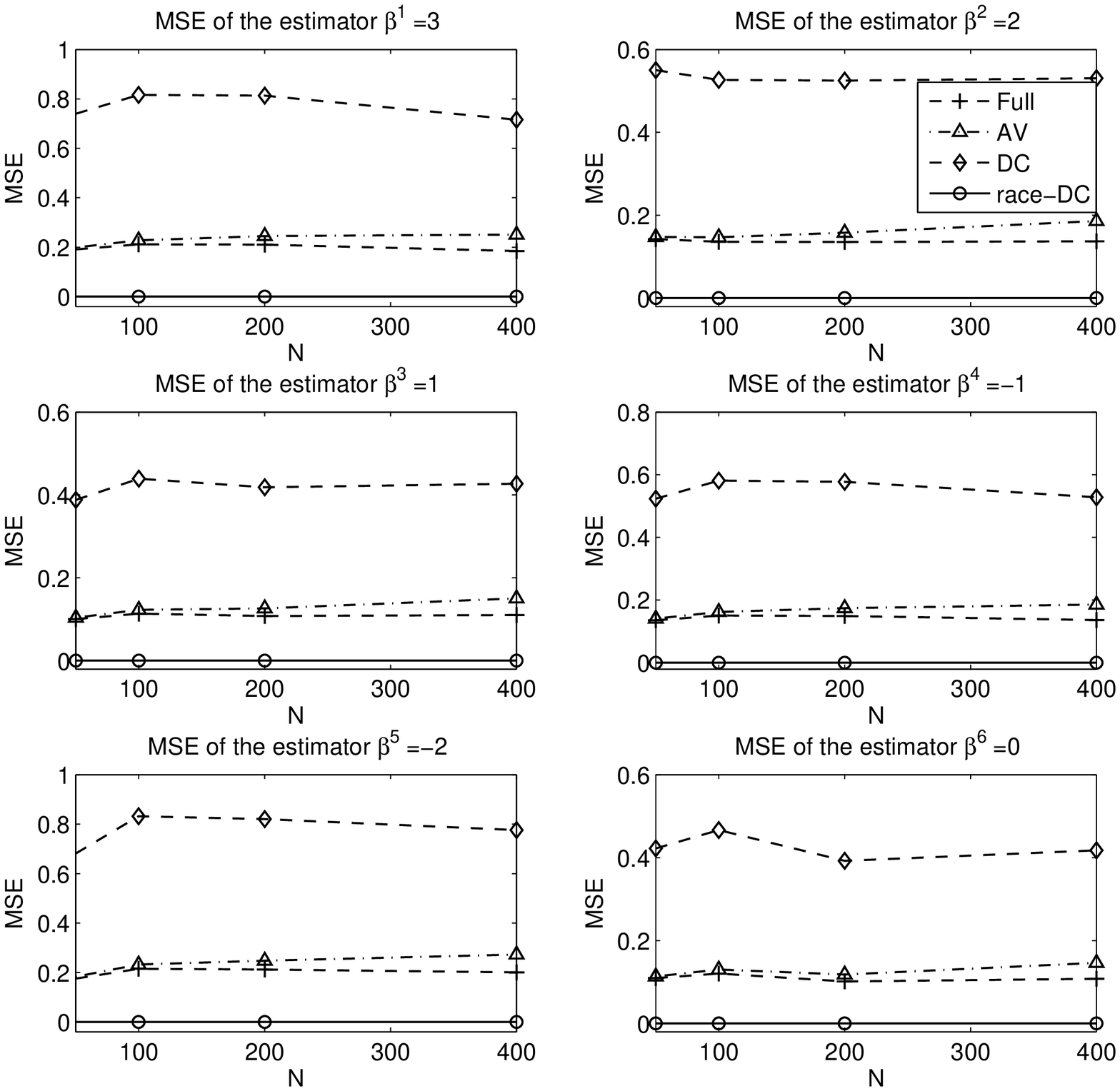}\vspace{-6ex}\\
	\caption{MSE of PCE  with $r=5$ in Experiment 3}\label{msepcet5}
\end{figure}

In this linear model, the correlation between the predictors is very high as well. This shows that we can use principle component-based methods to estimate parameters.
The DC expression of PCE is a little bit different from those of the Lasso and Ridge estimators. To make the DC algorithm applicable for PCE, we design the following algorithm:
\begin{itemize}
	\item[] {\it Step 1.} Each computer calculates the covariance matrix ${\bf X}_j{\bf X}_j^T$ and sends it to the central machine.
	\item[] {\it Step 2.} The central computer computes the spectral decomposition of $\sum_{j=1}{\bf X}_j{\bf X}_j^T$ as ${\bf P\Lambda P}^T$, where ${\bf P}$ is the matrix of eigenvectors and ${\bf\Lambda}={\rm diag}(\lambda_1,\cdots,\lambda_p)$ with eigenvalues $\lambda_1\geq\cdots\geq\lambda_p\geq 0$. Then, pass the first $r$ columns of ${\bf P}$ back to the computers distributed over the child nodes, where the number $r$ is determined by users.
	\item[] {\it Step 3.} Each computer establishes a new regressors as ${\bf Z}_{j,r}={\bf X}_j{\bf P}_r$ and again passes the new matrix ${\bf Z}_{j,r}{\bf Z}_{j,r}^T$ and the new vector ${\bf Z}_{j,r}{\bf y}_j$ to the central machine.
	\item[] {\it Step 4.} The central machine aggregates the compressed data and obtains the final DC estimator as
	$$\widehat{\beta}^{PCE}={\bf P}_r\left(\sum_{j=1}^N{\bf Z}^T_{j,r}{\bf Z}_{j,r}\right)^{-1}\sum_{j=1}^N{\bf Z}^T_{j,r}{\bf y}_j.$$	
\end{itemize}
Note that this algorithm requires each machine passing the compressed data twice. In the first time, as the spectral decomposition does not have a DC expression, to get a well-behaved estimate of eigenvector of a covariance matrix, each computer delivers the compressed sample covariance to central computer, then, the central computer constructs the spectral decomposition based on the aggregated covariances and sends the decomposed eigen-matrix back to each computer. In the second time, each computer again sends the compressed data to central computer to obtain the final estimates. The simple average of principle component estimator is defined by
\begin{equation*}
\overline\beta^{GDBP}=\frac1N\sum_{j=1}^N\widehat\beta_j^{DBP},
\end{equation*}
where $\widehat\beta_j^{DBP}$ is given by
\begin{eqnarray*}
	\widehat\beta^{DBP}_j=\widehat\beta_j^{PCE}+\frac1m {\bf M}_j {\bf X}_j^T({\bf y}_j-{\bf X}_j\widehat\beta_j^{PCE})
\end{eqnarray*}
with ${\bf M}_j=\left(\frac1m{\bf X}_j^T{\bf X}_j+k_1I\right)^{-1}$ and $\widehat\beta_j^{PCR}={\bf P}_r\left({\bf Z}^T_{j,r}{\bf Z}_{j,r}\right)^{-1}{\bf Z}^T_{j,r}{\bf y}_j$ be the PCE computed on ${\cal D}_j$. Similarly, the simple average method still needs twice communication between central computer and  computers on the child nodes.
\begin{center}
	Figures \ref{biaspcet4}-\ref{msepcet5} about here (attached at the end of the manuscript)
\end{center}
Figures \ref{biaspcet4}-\ref{msepcet5} report the empirical bias and MSE for the PCE-based estimators. Briefly speaking, the race-DC has a superior performance over all competitors. However, all the other methods, including the Full method, obtain the estimations with obvious deviation from the true parameters. Also note that the PCE under $r=5$ results in a reasonable small estimation bias compared to the case of $r=4$.

\subsection{Nonlinear model}
In this subsection, we examine the performance of the race-DC in nonlinear model. The sample is generated from the following nonlinear model:
\begin{equation*}
Y_i=(X_i^T\beta+2)^2+\varepsilon_i, \ i=1,\cdots,n,
\end{equation*}
where $\beta=(2,1,-2,0)^T$, $X_1,\cdots,X_n$, are independent and identically distributed as $N(0,\Sigma)$ with the $(i,j)$ element of $\Sigma$ equal to $0.5^{|i-j|}$, and $\varepsilon_1,\cdots,\varepsilon_n$, are independent and identically distributed as a standard normal distribution. In the simulation procedure, the sample size $n$ is chosen as 10000,  the number of batches of $N$ is set equal to $50,100,200,400$, respectively. We employ the algorithm proposed in Section \ref{nonlinear} to iterate out the final parameter estimation and we stop the algorithm unless the components $\beta^{(t+1)}$ and $\beta^{(t)}$ differ by the tolerance 1e-4. 

We compare our method with the simple average method, and the aggregated estimating equation (AEE) method. Actually, the AEE method is the DC method in nonlinear model deduced from the first-order Taylor expansion of the estimating equation, for the details see Lin and Xi (2011). In the iterative procedure, the initial value of $\beta$ is chosen as the estimator got by the data in the first batch.

\begin{figure}[H]
	\centering
	\includegraphics[width=6in]{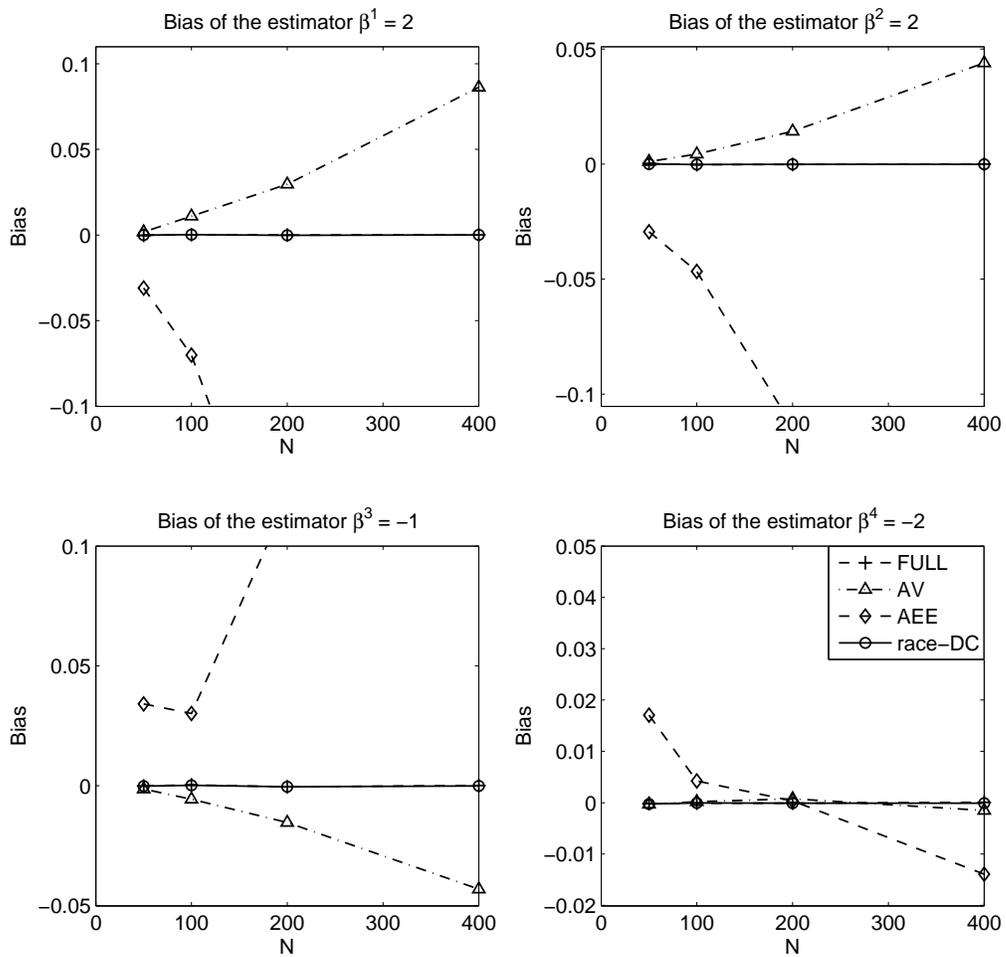}\vspace{-6ex}\\
	\caption{Bias for non-linear model in Experiment 4}\label{biasnonlmiid}
\end{figure}

\begin{figure}[H]
	\centering
	\includegraphics[width=6in]{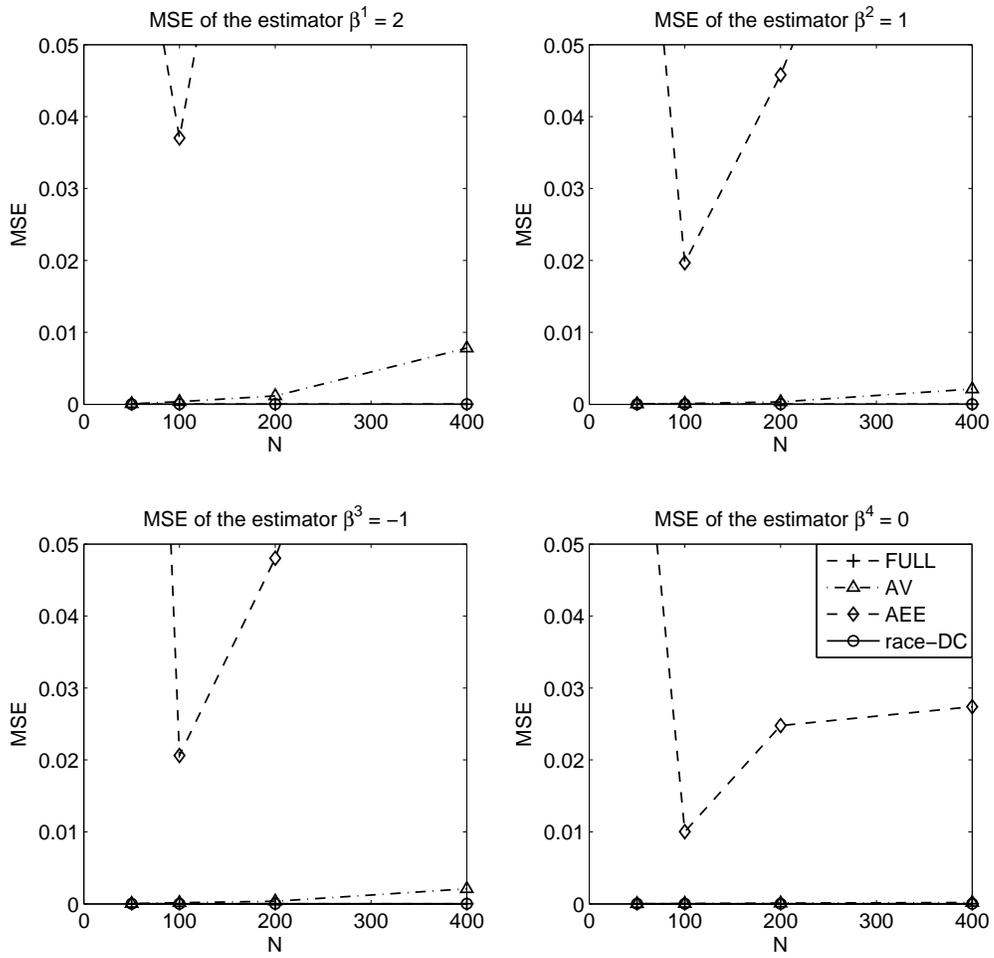}\vspace{-6ex}\\
	\caption{MSE for non-linear model in Experiment 4}\label{msenonlmiid}
\end{figure}

The median of the numbers of iterations in the race-DC algorithm is only 3. The numerical results are reported in Figures \ref{biasnonlmiid} and \ref{msenonlmiid}. We can obtain the following results:
\begin{enumerate}
	\item [(1)] Our race-DC behaves comparably well as the Full method, no matter how large the number of batches we choose;
\item [(2)] The simple average method has a nice performance for small $N$ but gets worse with the increase of $N$;
\item [(3)]
    The AEE method behaves very poor especially for large $N$. The reason may be that the aggregated Gram matrix sometimes could be degenerated under the chosen model setting.
\end{enumerate}

\begin{center}
	Figures \ref{biasnonlmiid}-\ref{msenonlmiid} about here (attached at the end of the manuscript)
\end{center}

Overall, by the four experiments respectively for linear model and nonlinear model, we conclude that the race-DC can lead to a good estimation no matter which model and bias estimator we chose. These can illustrate the theoretical conclusions that the regression composition in the race-DC procedure can completely correct the estimation bias in linear model and acceleratingly reduce the estimation bias in nonlinear model, regardless of the bias estimator we use.


\

\centerline{\Large \bf References}

\baselineskip=18pt

\begin{description}

\item Battey, H., Fan, J., Liu, H., Lu, J. and Zhu, Z. Distributed testing and estimation under sparse high dimensional models. {\it The Annals of statistics}, 46(3), 1352-1382, 2018.



\item Bradic, J., Fan, J. and Wang, W.
Penalized composite quasi-likelihood for ultrahigh
dimensional variable selection. {\it Journal of the Royal Statistical Society: Series B},
73(3), 325-349, 2011.


\item Chen, X., Liu, W. and Zhang, Y. Quantile regression under
memory constraint. {\it The Annals of Statistics}, 47(16), 3244-3273, 2019.

\item Chen, X. and Xie, M. G. A split-and-conquer approach for analysis of extraordinarily
large data. {\it Statistica Sinica}, 24(4), 1655-1684, 2014.

\item Chen, Y., Dong, G., Han, J., Pei, J., Wah, B. W. and Wang, J. Regression cubes with lossless compression and aggregation. {\it
IEEE Transaction on Knowledge and Data Engineering}, 18(12), 1585-1599, 2006.

\item Cheng, M. Y., Huang, T, Liu, P. and and Peng, H. Bias reduction for nonparametric and semiparametric regression models. {\it Statistica Sinica}, 28(4), 2749, 2018.

\item Dai, W. L., Tong, T. J. and Zhu, L. X. On the choice of difference sequence in a unified framework for variance estimation in nonparametric regression. {\it Statistical Science}, 32(3), 455-468, 2017.

\item Dai, W. L., Tong, T. J. and Genton, M. G. Optimal estimation of derivatives in nonparametric regression. {\it Journal of Machine Learning Research}, 17(1), 5700-5724, 2016.

\item Fard, M. M., Grinberg, Y., Pineau, J. and Precup, D. (2012). Compressed least-squares regression on sparse spaces. {\it Proceedings of the Twenty-Sixth AAAI Conference on Artificial Intelligence}, 1054-1060.





\item Javanmard, A. and Javanmard, A. Confidence intervals and hypothesis testing for
high-dimensional regression. {\it Journal of Machine Learning Research},
15(1), 2869-2909, 2014.


\item Jordan, M. I., Lee, J. D. and Yang, Y. Communication-efficient distributed
statistical inference. {\it Journal of the American Statistical Association}, 114(526), 668-681, 2019.




\item Kai, B, Li, R. and Zou, H. Local composite quantile regression smoothing: an efficient
and safe alterative to local polynomial regression. {\it Journal of the Royal Statistical Society: Series B}, 72(1), 49-69, 2010.

\item Kai, B, Li, R. and Zou, H. New efficient estimation and variable selection methods for
semiparametric varying-coefficient partially linear models. {\it The Annals of statistics}, 39(1), 305-332, 2011.


\item Lee, J. D., Liu, Q., Sun, Y. and Taylor, J. E. Communication-efficient sparse
regression. {\it Journal of Machine Learning Research}, 18(1), 115-144, 2017.

\item Li, R., Lin, D. K. and Li, B. Statistical inference in massive data sets. {\it Applied Stochastic Models in Business and Industry}, 29(5), 399-409, 2013.

\item Li, K. and Yang, J. Score-matching representative approach for big data
analysis with generalized linear models. arXiv preprint arXiv:1811.00462, 2018.

\item Lian, H., Zhao, K. and Lv, S. Projected spline estimation of the nonparametric function in high-dimensional partially linear models for massive data. {\it The Annals of Statistics}, 47(5), 2922-2949, 2018.

\item Lin, L. and Li, F. Stable and bias-corrected estimation for nonparametric
regression models. {\it Journal of Nonparametric Statistics},
20(4), 283-303, 2008.

\item Lin, L.,  Li, F., Wang, K. and Zhu, L. Composite estimation: An asymptotically weighted least squares approach. {\it Statistica Sinica}, 29, 1367-1393, 2019.

\item Lin, N. and Xi, R. Aggregated Estimating Equation Estimation. {\it Statistics and
Its Interface}, 4(1), 73-83, 2011.



\item Mcdonald, R., Mohri, M., Silberman, N., Walker, D. and Mann, G. S.
Efficient large-scale distributed training of conditional maximum entropy models. In {\it
Advances in Neural Information Processing Systems}, 1231-1239, 2009.

\item Meinshausen, N. and Buhlmann, P. Stability selection. {\it Journal of the Royal Statistical Society: Series B}, 72(4), 417-473, 2010.




\item Shah, R. D. and Samworth, R. J. Variable selection with error control: Another look at
stability selection. {\it Journal of the Royal Statistical Society: Series B}, 75(1), 55-80 2013.

\item Schifano, E. D., Wu, J., Wang, C., Yan, J., and Chen, M. H. Online updating of statistical inference in the big data setting. {\it Technometrics}, 58(3), 393-403, 2016.

\item Seber, G. A. F. and Wild, C. J. (2003). {\it Nonlinear regression}. John Wiley and
Sons, Inc.

\item Shi, C., Lu, W. and Song, R. A massive data framework for $M$-estimators with
cubic-rate. {\it Journal of the American Statistical Association}, 113(524), 1698-1709, 2018.


\item Sun, J., Gai, Y. and Lin, L. Weighted local linear composite quantile estimation for the case of general error distributions. {\it Journal of Statistical Planning and Inference}, 143(6), 1049-1063,2013.

\item Tibshirani, R. Regression shrinkage and selection via the lasso. {\it Journal of the Royal Statistical Society: Series B}, 58(1), 267-288, 1996.

\item Tong, T. and Wang, Y. Estimating residual variance in nonparametric regression using least
squares. {\it Biometrika}, 92(4), 821-830, 2005.


\item Volgushev, S., Chao, S. K. and Cheng, G. Distributed inference for quantile
regression processes. {\it The Annals of Statistics}, 47(3), 1634-1662, 2019.



\item  Wang, H., Yang, M. and Stufken, J.  Information-based optimal
subdata selection for big data linear regression. {\it Journal
of the American Statistical Association}, 114(525), 393-405, 2019.

\item Wang, J., Kolar, M., Srebro, N. and Zhang, T. Efficient distributed learning
with sparsity. In {\it Proceedings of the 34th International Conference on Machine Learning-Volume 70},  3636-3645, 2017.

\item Wang, W. W. and Lin, L. Derivative estimation based on difference sequence via
locally weighted least squares regression. {\it Journal of Machine Learning Research},
16(1), 2617-2641, 2015.

\item Wang, W., Yu, P., Lin, L. and Tong, T. Robust Estimation of Derivatives Using Locally Weighted Least Absolute Deviation Regression. {\it Journal of Machine Learning Research}, 20(60), 1-49, 2019.

\item Zinkevich, M., Weimer, M., Li, L., and Smola, A. J. Parallelized stochastic
gradient descent. In {\it Advances in Neural Information Processing Systems}, 2595-2603, 2010.

\item Zhang, Y., Duchi, J. and Wainwright, M. Divide and conquer kernel ridge
regression: A distributed algorithm with minimax optimal rates. {\it Journal of Machine Learning Research}, 16(1), 3299-3340, 2015.


\item Zhang Y., Duchi J. C. and  Wainwright, M. J. Communication-efficient algorithms for statistical optimization. {\it Advances in Neural Information Processing Systems}, 1502-1510, 2012.


\item Zhao, P. and Yu, B. On model selection consistency of Lasso. {\it Journal of Machine Learning Research}, 7, 2541-2563, 2006.

\item Zhao, T., Cheng, G. and Liu, H. A partially linear framework for massive
heterogeneous data. {\it The Annals of Statistics}, 44(4), 1400-1437, 2016.

\item Zou, H. and Yuan, M. Composite quantile regression and
the oracle model selection theory. {\it The Annals of Statistics}, 36(3), 1108-1126, 2008.

\end{description}

%
%
%

\end{document}